\documentclass[twocolumn,prl,preprintnumbers,superscriptaddress,landscape,nofootinbib]{revtex4}
\usepackage[utf8]{inputenc}
\usepackage{amsmath}
\usepackage{amsthm}
\usepackage{amssymb}
\usepackage{float}
\usepackage{braket}
\usepackage{graphicx}
\usepackage{subfigure}
\usepackage{url}
\usepackage{here}
\usepackage{natbib}
\usepackage{rotating}
\usepackage{CJKutf8}
\usepackage{bxcjkjatype}
\usepackage[colorlinks=true, pdfstartview=FitV, linkcolor=blue, citecolor=blue, urlcolor=blue]{hyperref}
\usepackage{orcidlink}
\usepackage{slashed}
\usepackage{multirow}
\usepackage[normalem]{ulem}
\graphicspath{{./figures/}}
\newcommand{\diff}{\mathrm{d}}

\newcommand{\p}{\partial}

\newcommand{\be}{\begin{equation}}      
\newcommand{\ee}{\end{equation}}      
\newcommand{\bea}{\begin{eqnarray}}      
\newcommand{\eea}{\end{eqnarray}}

\newcommand{\im}{\mathrm{i}}

\newcommand{\Tr}{\mathrm{Tr}}

\newcommand{\ctext}[1]{\raise0.2ex\hbox{\textcircled{\scriptsize{#1}}}}

\usepackage{tikz}
\usetikzlibrary{quantikz}

\theoremstyle{definition}

\theoremstyle{remark}

\usepackage[normalem]{ulem}

\begin{document}

\title{Beyond Energy: Teleporting Current, Charge, and More}
\author{Kazuki Ikeda (\begin{uCJK}池田一毅\end{uCJK})\orcidlink{0000-0003-3821-2669}}
\email[]{kazuki7131@gmail.com}
\affiliation{Co-design Center for Quantum Advantage $\&$ Center for Nuclear Theory, Department of Physics and Astronomy, Stony Brook University, Stony Brook, New York 11794-3800, USA}

\bibliographystyle{unsrt}

\begin{abstract}
As an homage to \textit{Quantum Energy Teleportation}, we generalize the idea to arbitrary physical observables, not limited to energy, and prove a rigorous upper bound on the activated (``teleported") quantity. The essence of this protocol is a quantum feedback control with respect to the entangled ground state of a quantum many-body system. To demonstrate the concept, we explore a (1+1)-dimensional chiral Dirac system and execute the protocol for the electric current and charge. One of the most significant results is the creation of long-range correlations across the system after applying control operations only to one local site. Consequently but surprisingly, the induced charge susceptibility fully reconstructs the phase diagram, despite the model initially having no charge. Moreover, we find an activation of novel chiral dynamics induced by feedback control operations, which can be experimentally confirmed using trapped ions and neutral atoms.
\end{abstract}

\maketitle

\emph{Introduction}.---
Quantum feedback control protocols utilize efficient and innovative methods to manipulate quantum systems and process quantum information \cite{2017PhR...679....1Z}. These methods significantly contribute to understanding thermodynamics at quantum scales from the perspective of information theory. Their scope extends beyond physics into areas such as quantum biology \cite{PhysRevA.106.032218} and chemistry \cite{1997CPL...280..151B}. Additionally, they are integral to advancing quantum technologies by enhancing the stability, accuracy, and reliability of quantum operations \cite{2023NatCo..14.4721G}.

Quantum Energy Teleportation (QET) \cite{PhysRevD.78.045006,HOTTA20085671,2010PhLA..374.3416H,2011arXiv1101.3954H}, recently experimentally verified \cite{Ikeda:2023uni,Ikeda_Quantum_Energy_Teleportation_2023,Ikeda:2024hbi,PhysRevLett.130.110801}, enables energy transfer between distant quantum parties using local operations and classical communications (LOCCs). Leveraging quantum entanglement and feedback control, QET is gaining attention in condensed matter, high energy physics, and quantum communication~\cite{Ikeda:2023ljh,PhysRevD.107.L071502,https://doi.org/10.1049/qtc2.12090,10.1116/5.0164999,Ikeda:2023yhm,Wang:2024yqn,Itoh:2023pmj}.

\begin{figure*}
\begin{minipage}{0.32\linewidth}
    \centering
    \includegraphics[width=\linewidth]{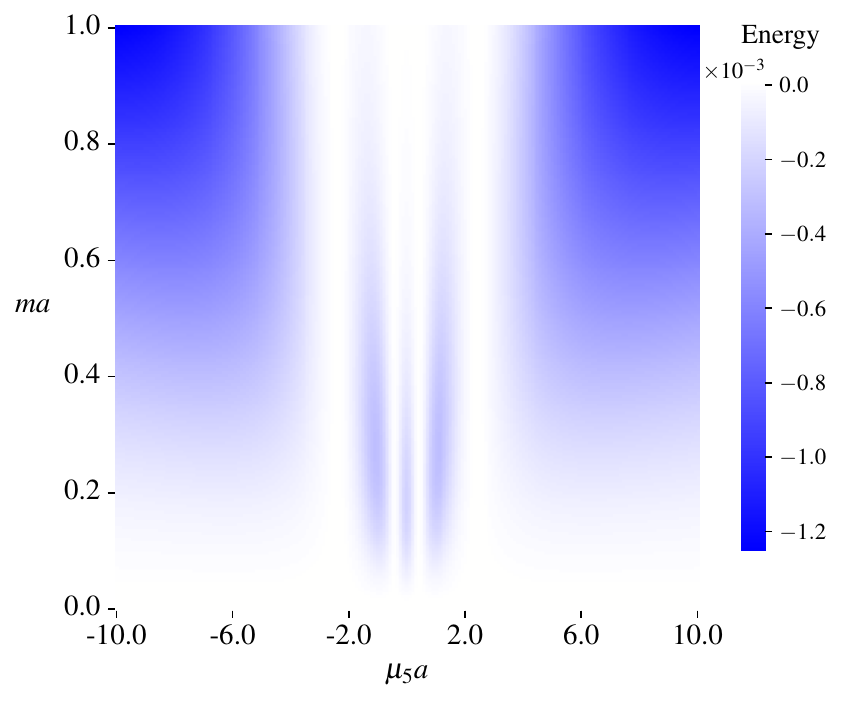}
\end{minipage}
\begin{minipage}{0.32\linewidth}
    \centering
    \includegraphics[width=\linewidth]{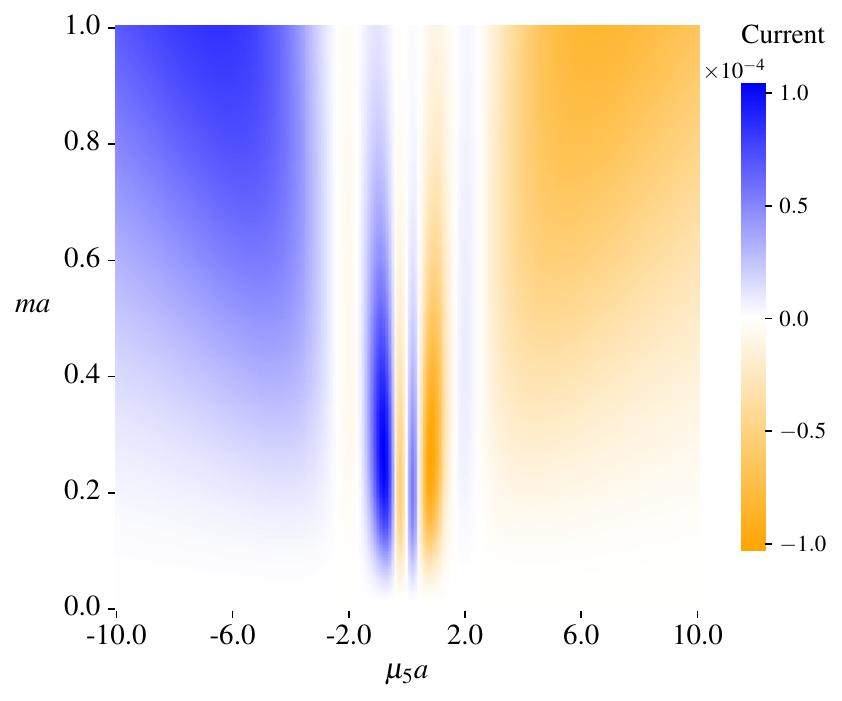}
\end{minipage}
\begin{minipage}{0.32\linewidth}
    \centering
    \includegraphics[width=\linewidth]{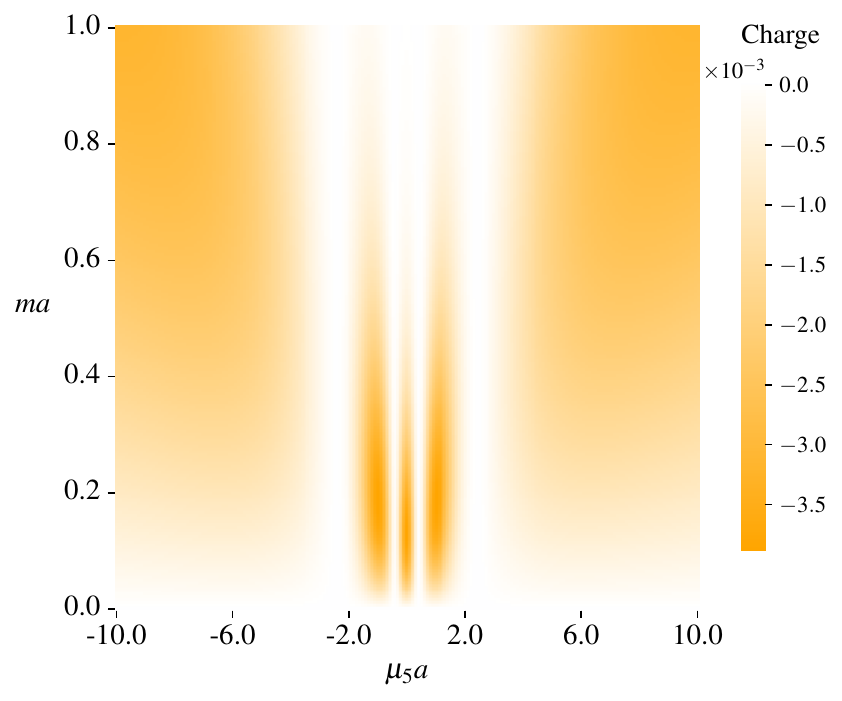}
\end{minipage}
    \caption{From left to right: Energy $\langle\Delta E_{n_B}\rangle$, electric current $\langle\Delta J^1_{n_B}\rangle$ and electric charge $\langle\Delta J^0_{n_B}\rangle$, computed with the lattice of $N=8$ qubits under the OBC.}
    \label{fig:teleportation}
\end{figure*}

We aim to answer the following open questions: 
\begin{enumerate}
\item\label{prob:1} Can QET be generalized to an arbitrary quantity, not only energy? (solved in eq.~\eqref{eq:delta}, Fig.~\ref{fig:teleportation} \& \hyperref[sec:step2]{SM})
\item\label{prob:2} How much can it be activated? (solved in eq.~\eqref{eq:formula})
\item\label{prob:3} How is it related to entanglement?(Figs.~\ref{fig:entropy_charge},\ref{fig:correlation_function}\&\hyperref[sec:bound]{SM})
\item\label{prob:4} What are new applications/phenomena? (Fig.~\ref{fig:wave})
\end{enumerate}
We develop the protocol to encompass quantities beyond energy, namely activating general physical quantities such as charge, current, and even random observables. Teleportation of current and charge should be realised experimentally as a natural result of QET. The core of the protocol is applying quantum feedback control to the ground state of an entangled quantum many-body system. This approach aims to expand quantum information science by demonstrating the teleportation of various physical quantities through entangled states.

To elucidate the underlying principles, we investigate the protocol in the following (1+1)-dimensional chiral Dirac system, whose Lagrangian density is
\begin{align}
\label{eq:model}
\mathcal{L}=\bar\psi\left(\gamma^\mu\partial-m\right)\psi-\mu_5\bar\psi\gamma^1\psi,
\end{align}
where $\gamma^\mu$ are Dirac matrices. Non-zero $\mu_5$ induces chiral imbalance between the left- and right-handed fermions. This is a simplified model of a quantum field theory with minimal parameters. For spin chain Hamiltonian~\eqref{eq:Ham_pbc}, see the Supplemental Material (\hyperref[sec:Ham]{SM}), where the work is also extended to the (1+1)-dimensional QED, which serves as an ideal benchmark for quantum simulations with quantum and classical hardware~\cite{Klco:2018kyo,Farrell:2023fgd,Farrell:2024fit,Butt:2019uul,Magnifico:2019kyj,Shaw:2020udc,Florio:2024aix,2023arXiv230111991F, Qian:2024xnr,PhysRevLett.122.050403,2020arXiv200100485C,Czajka2022,PhysRevResearch.2.023342}.

\emph{Chiral Maxwell Demon}.--- Quantum feedback control protocols can be regarded as a modern, practical embodiment of Maxwell's demon \cite{PhysRevA.56.3374,szilard1929entropieverminderung,PhysRevLett.100.080403,PhysRevLett.102.250602,PhysRevLett.106.189901,PhysRevLett.104.090602,PhysRevLett.108.030601,2010NatPh...6..988T,2019NatPh..15..660R}, applying the principles of measurement and information-based control to manage and manipulate quantum systems. This connection bridges thermodynamics with applications in quantum technologies, both theoretically and experimentally.

The chirality of a Dirac fermion in the model~\eqref{eq:model} can be measured by the projection operator:
\begin{equation}
P(b)= \frac{1}{2}\left(1 - (-1)^b \gamma_{5}\right),
\end{equation}
where $b = 0$ for the right-handed mode and $b = 1$ for the left-handed mode. The chiral Maxwell Demon (CMD) is characterized by its ability to measure the chirality of fermions and execute a control protocol. Conducting the chirality measurements, the CMD records the outcomes as a bit-string $b_0 b_1 \cdots b_k$.

In the protocol, the CMD involves spacially separated Alice (detector) and Bob (controller). Let $\ket{gs}$ be the ground state of the system and $\rho_0=\ket{gs}\bra{gs}$ the corresponding density matrix. Alice measures her local qubit of $\ket{gs}$ by $P_{n_A}(b)$ and obtains either $b=0$ or $1$. Repeating the measurements, Alice creates a mixed state $\rho_\text{post}=\sum_bP_{n_A}(b)\rho_0P_{n_A}(b)$, and therefore excites her local energy: $\langle\Delta E_{n_A}\rangle=\Tr[\rho_\text{post}H_{n_A}]-\Tr[\rho_0 H_{n_A}]>0$. 

Receiving Alice's outcome $b$, Bob applies a control operation $U_{n_B}(b)=e^{-b\gamma^1_{n_B}}$ to his qubit. Then the CMD statistically obtains the feedback-controlled (FC) state:
\begin{equation}
\label{eq:rho_QET}
    \rho_\text{FC}=\sum_{b\in\{0,1\}}U_{n_B}(b)P_{n_A}(b)\rho_0P_{n_A}(b)U^\dagger_{n_B}(b). 
\end{equation}

The CMD is fully consistent with information thermodynamics. Let $X$ be the system after the measurement and $X'$ be the evolution of $X$ after feedback. The second law of thermodynamics is equivalent to \cite{PhysRevLett.100.080403,Parrondo2015}
\begin{equation}
\label{eq:2ndlaw}
    S(X')-S(X)\ge I(X':A)-I(X:A),
\end{equation}
where $S(\cdot)$ is the von Neumann entropy of $\cdot\in\{X,X'\}$ and $I(\cdot:A)$ is the mutual information between  $\cdot$ and Alice's memory $A$. In our case, both L.H.S and R.H.S of \eqref{eq:2ndlaw} are negative, which confirms that $X$ receives a feedback from Alice. This clarifies the relation between the Demon and the QET protocol. See also \hyperref[sec:demon]{SM}. 

\emph{The upper bound of teleportation}.---
Applying $\rho_\text{FC}$ to an arbitrary observable $\mathcal{O}$ would activate some amount, which we call \textit{teleportation} of $\mathcal{O}$ following QET. Here we provide a general concise answer \eqref{eq:formula} to \hyperref[prob:2]{Question \ref{prob:2}}: the upper bound of the teleported amount, which has remained an open question, particularly in the context of QET. 

We define the difference between the two expectation values of $\mathcal{O}$ evaluated by two density operators $\rho,\sigma$: 
\begin{equation}
\label{eq:delta}
    \Delta \mathcal{O}=\Tr[\rho \mathcal{O}]-\Tr[\sigma \mathcal{O}]. 
\end{equation}
Then we note that the upper bound of the trace norm of $\Delta \mathcal{O}$ is related to the fidelity in such a way that 
\begin{equation}
    \|\Delta \mathcal{O}\|\le \|\rho-\sigma\|\|\mathcal{O}\|\le 2\sqrt{1-F(\rho,\sigma)}\|\mathcal{O}\|,
\end{equation}
where $F$ is the fidelity between two density matrices $\rho,\sigma$:
\begin{equation}
\label{eq:fidelity}
F(\rho,\sigma)=\left(\Tr\sqrt{\sqrt{\rho}\sigma\sqrt{\rho}}\right)^2. 
\end{equation}
Since $|\langle\Delta \mathcal{O}\rangle|=|\Tr[\Delta \mathcal{O}]|\le\|\Delta \mathcal{O}\|$, the following inequality holds and gives the upper bound of teleportaiton:  
\begin{equation}
\label{eq:formula}
|\langle\Delta\mathcal{O}\rangle|\le 2\sqrt{1-F(\rho,\sigma)}\|\mathcal{O}\|. 
\end{equation}

\emph{Quantum Current and Charge Teleportation}.---
Here we focus on the electric current and the electric charge of the model~\eqref{eq:model}. Our operator is defined by 
\begin{equation}
    J^\mu(t,x)=\bar\psi(t,x)\gamma^\mu\psi(t,x),~\mu=0,1.
\end{equation}

We map the model to a staggered
lattice~\cite{Kogut:1974ag,Susskind:1976jm,Jordan:1928wi} having $N$ qubits and, throughout this work, we put Bob at the center $(n_B=N/2)$ and Alice at $n_A=n_B-N/2$. We the open boundary conditions (OBC) in the main text and the periodic boundary conditions (PBC) in \hyperref[sec:Ham]{SM}. For the given setup, \textit{Quantum Current and Charge Teleportation} are defined by applying the density operator~\eqref{eq:rho_QET} to the charge $J^0_{n_B}$ and the current $J^1_{n_B}$ at Bob's local coordinate. In Fig.~\ref{fig:teleportation}, we present energy $\langle\Delta E_{n_B}\rangle$, electric current $\langle\Delta J^1_{n_B}\rangle$, and electric charge $\langle\Delta J^0_{n_B}\rangle$ evaluated by eq.~\eqref{eq:rho_QET}. In contrast to energy and charge, the current flips the sign depending on chiral chemical potential $\mu_5$ since the direction of the current is determined by the imbalance between the left and right modes.  

\begin{figure}[H]
    \centering
    \includegraphics[width=0.8\linewidth]{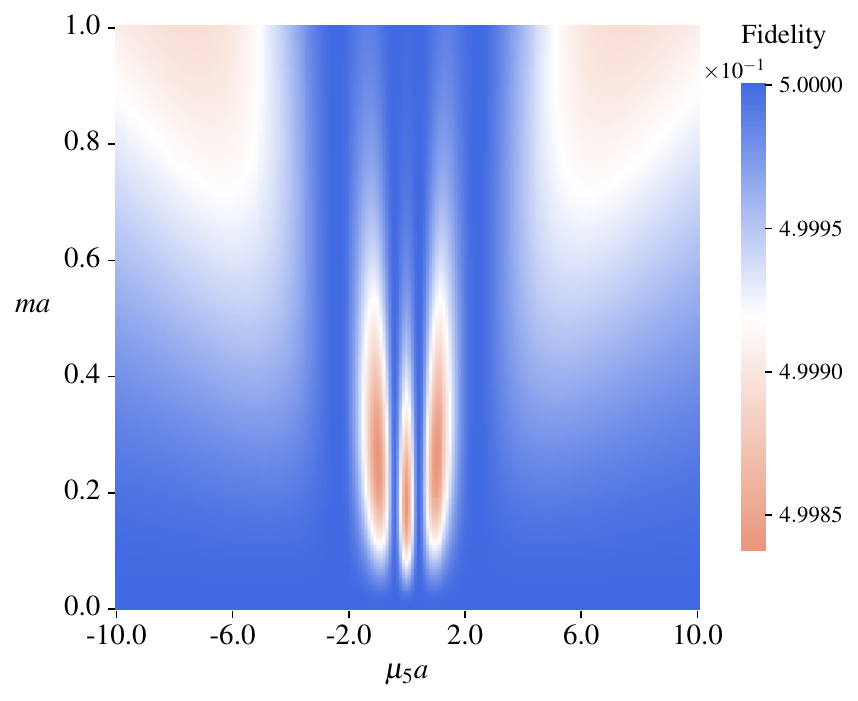}
    \caption{The fidelity between $\rho_0=\ket{gs}\bra{gs}$ and $\rho_\text{FC}$~\eqref{eq:rho_QET}. }
    \label{fig:fidelity}
\end{figure}

\begin{figure}[H]
\begin{minipage}{0.49\linewidth}
    \centering
    \includegraphics[width=\linewidth]{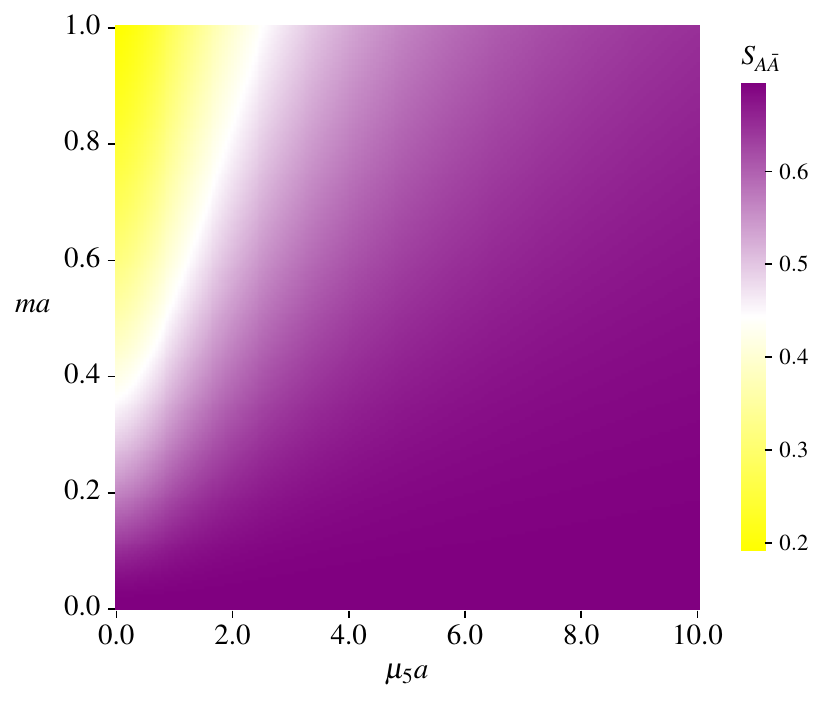}
\end{minipage}
\begin{minipage}{0.49\linewidth}
    \centering
    \includegraphics[width=\linewidth]{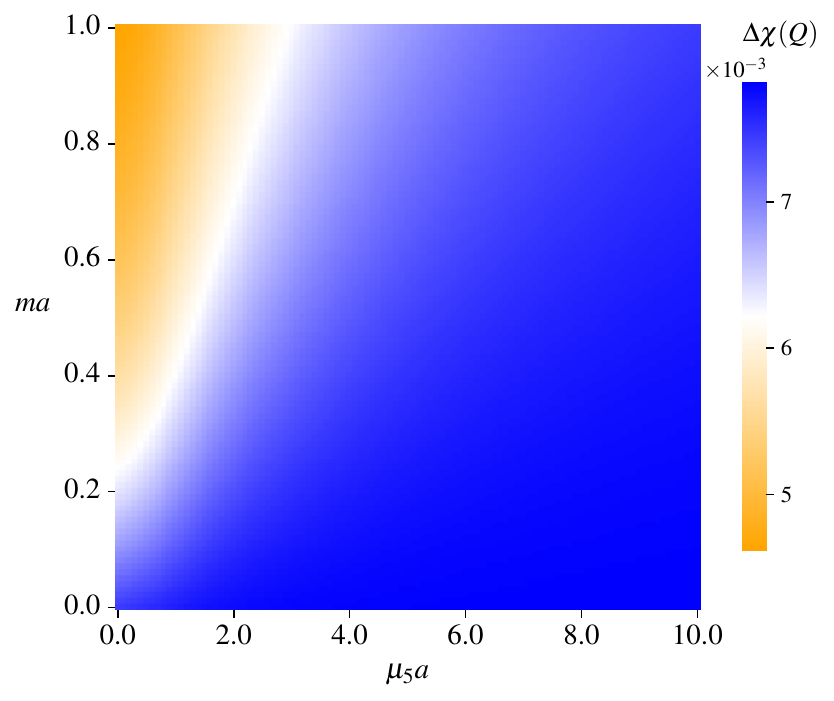}
\end{minipage}
\caption{Comparison between the ground state entanglement entropy $S_{A\bar{A}}(\rho_0)$ (left) and the charge susceptibility $\Delta\chi(Q)=\chi(Q,\rho_\text{FC})$ induced by the feedback-control (right).}
    \label{fig:entropy_charge}
\end{figure}

\begin{figure}
    \centering
    \includegraphics[width=\linewidth]{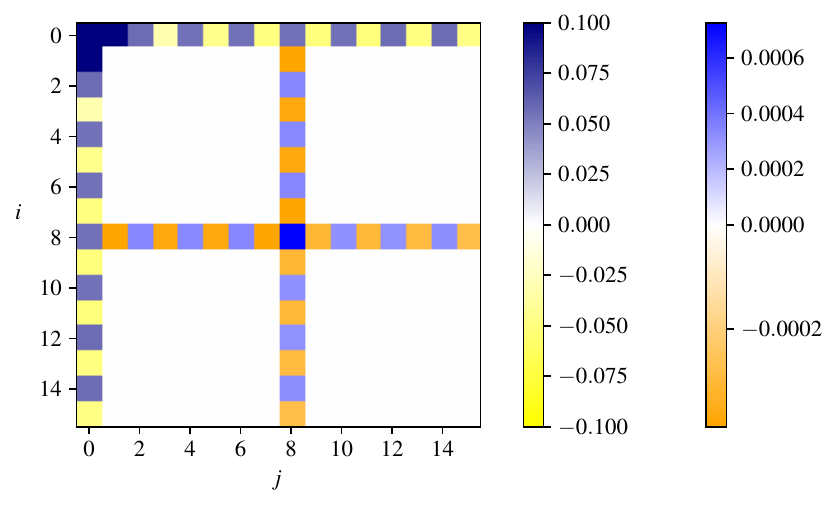}
    \caption{The spacial charge-charge correlation function $\Delta\Pi^{00}_{ij}$ at $t=0$. Parameters: $ma=0.1,\mu_5a=1,N=16$.}
    \label{fig:correlation_function}
\end{figure}

\begin{figure*}
    \centering
    \subfigure[~$J^{\mu}_\text{post}(t,x)$]{\includegraphics[width=0.42\textwidth]{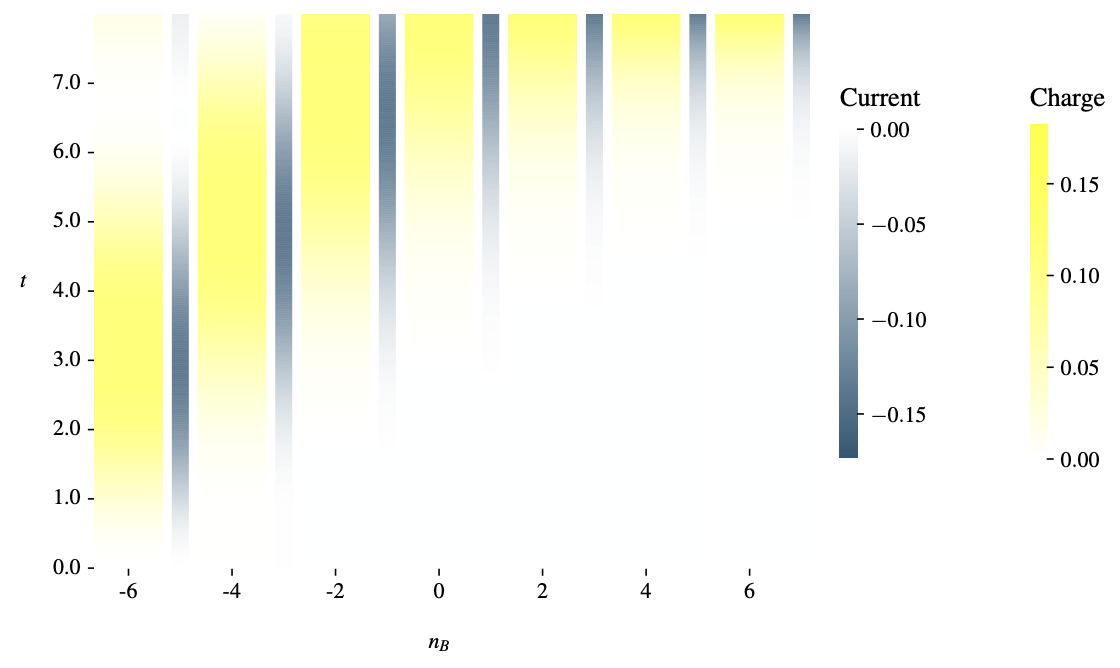}}
    \hskip1.5cm
    \centering
    \subfigure[~$J^{\mu}_\text{post}(t,x)-J^{\mu}_\text{FC}(t,x)$]
    {\includegraphics[width=0.42\textwidth]{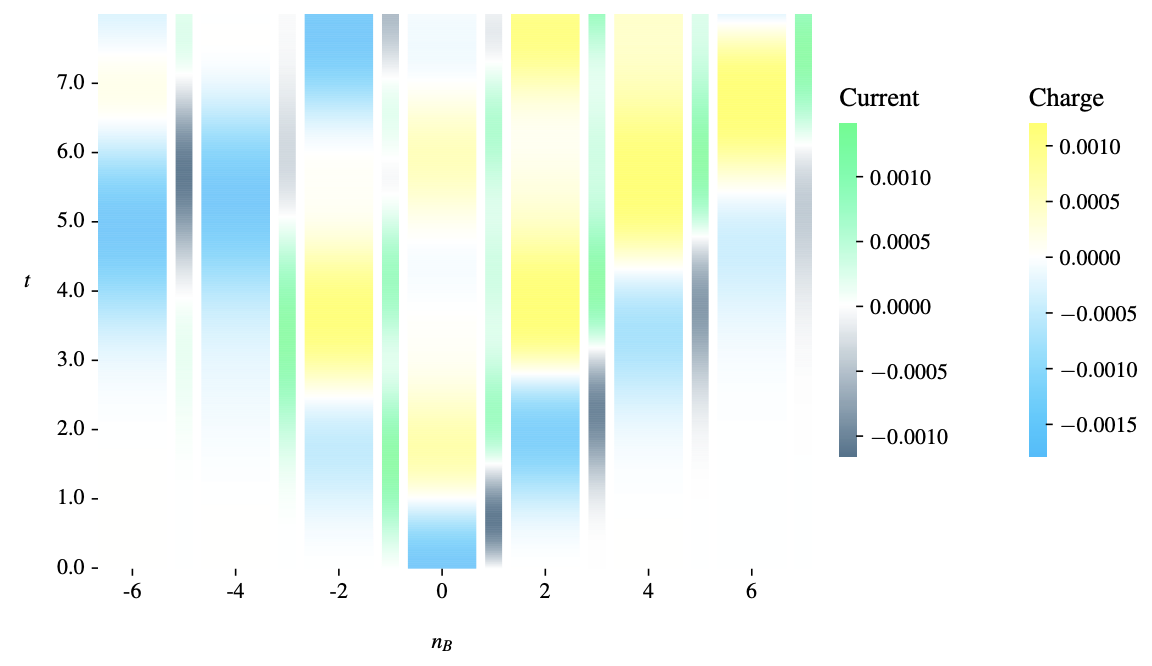}}
    \caption{Real-time evolution of current and charge  after Alice's measurement and Bob's controlled operation, viewed from Bob's coordinate $n_B$ centered to the $N=16$ open boundary lattice. Parameters: $ma=0.1,\mu_5=1$. }
    \label{fig:wave}
\end{figure*}

To provide an intuitive grasp of the inequality~\eqref{eq:formula}, we present the fidelity~\eqref{eq:fidelity} in Fig.~\ref{fig:fidelity}. The significant relationship with the teleported quantities is apparent in Fig.~\ref{fig:teleportation}. Notably, the peaks in energy, electric current, and electric charge align with the minimal values of the fidelity. Moreover, this relationship underscores the global trend that lower fidelity results in more efficient teleportation, whereas higher fidelity leads to less teleportation, which gives another confirmation of eq.~\eqref{eq:formula}. See also \hyperref[sec:bound]{SM}. 

\emph{Induced susceptibilities and quantum correlations}.---
For a given observable $\mathcal{O}$ and a density operator $\rho$, the susceptibility $\chi\left(\mathcal{O},\rho\right)$ is defined by 
\begin{align}
   \chi\left(\mathcal{O},\rho\right)&=\Tr[\rho \mathcal{O}^2]-\Tr[\rho\mathcal{O}]^2.
\end{align}
In~\cite{PhysRevD.103.L071502,PhysRevD.108.L091501} we argued that some susceptibilities are good order parameters of the phase transition in the (1+1)-d QED. It turns out that the susceptibility is a powerful tool to investigate \hyperref[prob:3]{Question \ref{prob:3}}. To explore the change in the susceptibility following the implementation of feedback-control, we consider 
\begin{equation}
\Delta\chi(\mathcal{O})=\chi\left(\mathcal{O},\rho_\text{FC}\right)-\chi\left(\mathcal{O},\rho_0\right).
\end{equation}
In what follows, we discuss the charge and denote the average charge by $Q=\frac{1}{N}\sum_{n}J^0_n$. We also explored the electric susceptibility in \hyperref[sec:electric]{SM}. It should be noted that the charge susceptibility is exactly 0 with respect to the ground state of the model $(\chi(Q,\rho_0)=0)$). However $\chi(Q,\rho_\text{FC})$ is no longer 0, since the measurements and controlled operations excite the energy level, by which the state could be charged. In Fig.~\ref{fig:entropy_charge} (see Fig.~\ref{fig:PBC_charge_susceptibility} in \hyperref[sec:PBC_susceptibility]{SM} for PBC), we present the change in the charge susceptibility ($N=8$):
\begin{equation}
\label{eq:delta_susceptibility}
    \Delta\chi(Q)=\frac{1}{N^2}\sum_{ij}\Tr\left[\rho_\text{FC}J^0_iJ^0_j\right]
\end{equation}

Remarkably the diagram of $\Delta\chi(Q)$ captures the significant qualitative property of the phase diagram
drawn by the ground state entanglement entropy $S_{A\bar{A}}(\rho_0)$ between Alice $A$ and its complement $\bar{A}$. This agreement is extremely nontrivial because of the following reasons. First of all, $\Delta\chi(Q)$ is entirely determined by $\rho_\text{FC}$ since $\chi(Q,\rho_0)=0$. However, as confirmed in Fig.~\ref{fig:teleportation}, teleportation is apparently not directly related to entanglement, although its upper bound is related to the entanglement entropy and fidelity, as discussed in \hyperref[sec:bound]{SM}. In fact, $\Delta\chi(Q)$ holds a large value even at a point where teleportation does not occur at all. 

This counterintuitive feature suggests the emergence of long-range correlations between two sites, although Alice and Bob's operations are local and LOCCs do not enhance entanglement entropy.

To understand this phenomenon better, we explore the current-current correlation function~\cite{Barata:2024bzk,code}:
\begin{align}
\label{eq:QQ-correlation}
\Pi^{\mu\nu}(\rho,t, x _i,x_j)&=\Tr[\rho J^\mu(t,x_i)J^\nu(0,x_j)]. 
\end{align}
Similar to eq.~\eqref{eq:delta}, we consider 
\begin{equation}
\label{eq:diff_QQ-correlation}
    \Delta\Pi^{\mu\nu}_{ij}(t)=\Pi^{\mu\nu}(\rho_\text{FC},t, x _i,x_j)-\Pi^{\mu\nu}(\rho_0,t, x _i,x_j).
\end{equation}
In particular $\Delta\Pi^{00}_{ij}(t=0)$ is the spatial charge-charge correlation function and allows us to examine each term in the susceptibility~\eqref{eq:delta_susceptibility}. A result is plotted in Fig.~\ref{fig:correlation_function} and more details including dynamics at $t>0$ are presented in \hyperref[sec:QQ]{SM} and related to the operator growth~\cite{Roberts:2014isa,Hosur:2015ylk,PRXQuantum.5.010201}. The results would give us a new insight into the quantum correlation induced by LOCCs and quantum feedback-controls. The nontrivial aspects of the results are summarized below:
\begin{itemize}
    \item $\Delta\Pi^{00}_{ij}\neq0$ when $i$ or $j$ is $n_A$ or $n_B$. This clearly indicates that some correlations are created by quantum feedback-control. It is highly nontrivial that both of Alice and Bob's local operations induce long-range correlations across the system.  
    \item $\Delta\Pi^{00}_{i,2j}+\Delta\Pi^{00}_{i,2j+1}>0$ when $i$ is $n_A$ or $n_B$. This was confirmed numerically. For the ground state of the system, the neighboring sits have opposite charges due to the staggered lattice formulation and the total charge is conserved. However, this is not the case for $\rho_\text{FC}$, which has a charge. 
    \item Consequently, the charge susceptibility is positive: $\Delta\chi(Q)\propto\sum_{ij}\Delta\Pi^{00}_{ij}>0$, which agrees with Fig.~\ref{fig:entropy_charge}. 
\end{itemize}

We explored the quantum resources for QET in \cite{Ikeda:2024hbi} and argued that some quantum correlations, called quantum discord~\cite{PhysRevLett.88.017901}, are not completely destroyed during QET, although Alice's measurement destroys entanglement. Our new results revealed the emergence of correlations in the mixed state created by the feedback-control. Purification of $\rho_\text{FC}$ would be helpful to solve this puzzle.

\emph{Yet another chiral dynamics - For experiment}.---
Here we consider the dynamics after Alice's measurement and Bob's control operation, to explore \hyperref[prob:4]{Question \ref{prob:4}} more. Using real-time evolution $\rho(t)=e^{-itH}\rho e^{itH}$ of a density matrix $\rho$, we evaluate the real-time evolution of charge and current as follows:
\begin{align}
    J^\mu_{\text{post}}(t,x)&=\frac{\Tr[\rho_\text{post}(t)J^\mu(0,x))]}{\Tr[\rho_0J^\mu(0,x))]},\\
    J^\mu_{\text{FC}}(t,x)&=\frac{\Tr[\rho_\text{FC}(t)J^\mu(0,x))]}{\Tr[\rho_0J^\mu(0,x))]}. 
\end{align}
We assume that Alice and Bob communicate instantly and that $\rho_\text{post}$ and $\rho_\text{FC}$ are created without duration. 

Let us first remind that Alice measures the ground state $\rho_0$ in $X$-basis, which flips the local spin and induces a charge. Therefore the post-measurement state $\rho_\text{post}$ is no longer the ground state and obeys the quench dynamics for $t>0$. Consequently, $J^\mu_{\text{post}}(t,x)$ propagates on a light-cone for a small mass as exhibited in Fig.~\ref{fig:wave}~{(a)}, where Alice is located at $n_B-N/2$.

The same things can be said to the time-evolution of the feedback-controlled state $\rho_\text{FC}$, since Bob applies $U_{n_B}(b)$ to $\rho_\text{post}$, which becomes also excited. However it should be noted that the rotation angle $\phi$ used for Bob's operation is very small and therefore $\rho_\text{post}$ and $\rho_\text{FC}$ are almost identical. In fact, one can find that $J^{\mu}_\text{FC}(t,x)$ shows a similar behavior as $J^{\mu}_\text{post}(t,x)$. However this is not the end of the story. When we explore the difference 
\begin{equation}\label{eq:wave}
    J^\mu_{\text{post}}(t,x)-J^\mu_{\text{FC}}(t,x),
\end{equation}
we can find the propagation of the waves emitted from Bob's coordinate as illustrated in Fig.~\ref{fig:wave}~{(b)}. The upper bound of the amplitude~\eqref{eq:wave} is related to the fidelity $F(\rho_\text{post},\rho_\text{FC})$ as given by eq.~\eqref{eq:formula} and that the current $J^\mu(t,x)=\Tr[\rho(t)J^\mu(0,x)]$ should respect the conservation law $\partial_\mu J^\mu=0$ for $\rho=\rho_0,\rho_\text{post},\rho_\text{FC}$.  

Although our system lacks an electric field and thus does not have chiral anomaly, the light-conic wave propagation (from both Alice and Bob's sites) reminds us of chiral magnetic waves~\cite{PhysRevC.89.044909,PhysRevLett.107.052303,Kharzeev:2010gd,STAR:2015wza}. For a larger mass, we can also confirm nonlinear waves, called ``Thumper"~\cite{PhysRevD.108.074001}, consisting of a rapidly oscillating vector current (axial charge) and a slowly oscillating electric charge (axial current). 

Asymmetry in the waves (Fig.~\ref{fig:wave}~(b)) can be described as follows. Because of the configurations in OBC, Alice's wave propagates from left to right, whereas Bob's wave propagates from the center to the both directions. Hence Alice's wave and Bob's left-going wave collide at some point, while Bob's right-going wave will not collide to Alice's wave. Besides, the system's parity symmetry is initially broken due to the chiral chemical potential. To observe waves without the effect of collision between Alice and Bob's waves, we need to take a larger lattice having a long quantum coherence time. 

Trapped ions provide an excellent platform for the experimental verification of our results~\cite{Blatt2012,RevModPhys.93.025001}. Quantinuum recently announced a trapped-ion quantum computer with 56 qubits~\cite{2024arXiv240602501D}. Trapped ions feature long coherence times, preserving quantum states longer, crucial for observing the propagation of the waves in Fig.~\ref{fig:wave}. Laser-based precise manipulation of these ions allows for accurate quantum gate control, essential for error correction. Technologies like the quantum CCD (QCCD) enhance scalability, aiding in the creation of larger quantum processors. High-fidelity gates and measurements further reduce operational errors, improving the reliability of quantum feedback control and time evolution. 

Rydberg atoms also provide an intriguing platform~\cite{rydberg1, rydberg2, rydberg3,2024arXiv240404411B}. Electrons in these large orbits can be manipulated through light-matter interactions and possess substantial dipole moments linking neighboring atoms. Mutual interactions within a Rydberg atom ensemble ensure only one atom is excited at a time. When dipole interactions are much smaller than the energy level differences between atoms, they can be treated perturbatively. This platform of neutral atoms offers excellent scalability, enabling the resolution of large-scale problems~\cite{Ebadi:2022oxd,2022Natur.604..451B,Bluvstein:2023zmt}.

\emph{Discussion - Toward Quantum Battery}.---
The main contributions of our work to QET, and more broadly to quantum feedback and control in quantum many-body systems, can be summarized as follows:
\begin{description}
    \item[{Answer to \hyperref[prob:2]{Question \ref{prob:1}}}]: Adapting the protocol to encompass arbitrary observables. 
     \item[{Answer to \hyperref[prob:2]{Question \ref{prob:2}}}]: Asserting the most general upper bound of teleportation~\eqref{eq:formula}. Generally this could be a stronger upper bound than the known bound given by the entanglement entropy $S_{A\bar{A}}(\rho_0)$ \cite{PhysRevA.87.032313}, which can exceed 1. The formula is further generalized in \hyperref[sec:bound]{SM}.
    \item[{Answer to \hyperref[prob:3]{Question \ref{prob:3}}}]: Ascertaining the activation of global charge-charge correlations (Figs. \ref{fig:entropy_charge},\ref{fig:correlation_function}), which cannot be achieved in the standard physics framework of the unitary evolution, as the model \eqref{eq:model} is charge neutral.
    \item[{Answer to \hyperref[prob:4]{Question \ref{prob:4}}}]: Activating new chiral dynamics following the feedback (Fig.~\ref{fig:wave}). 
\end{description}

We emphasize that our findings are expected to appear in various physical systems. A mid-circuit measurement of the ground state raises its energy, inducing local currents and charges. Control operations can then instantaneously activate energy, currents, and charges at distant locations, triggering novel dynamic phenomena.

To provide alternative answer to \hyperref[prob:4]{Question \ref{prob:4}}, let us conclude this work by extending our discussion to quantum batteries (QB), which are energy storage devices utilizing many-body quantum systems.~\cite{PhysRevE.87.042123,Binder_2015,PhysRevLett.118.150601,Campaioli:2023ndh,PhysRevLett.128.140501,PhysRevA.108.042618,PhysRevLett.125.236402}. First, it would be natural to extend the idea of teleportation to the highest energy state $\ket{E_{\max}}$ instead of the ground state $\ket{gs}$, since $\ket{E_{\max}}$ is the ground state of the sign-flipped Hamiltonian $-H$. If Alice and Bob perform the QET protocol to $\rho_{\max}=\ket{E_{\max}}\bra{E_{\max}}$ and obtain $\rho_\text{QB}$ by~\eqref{eq:rho_QET}, then Alice extracts energy and Bob can charge energy beyond his maximal local energy:
\begin{equation}
\Tr[\rho_\text{QB}H_{n_B}]>\Tr[\rho_{\max}H_{n_B}].    
\end{equation}
In this sense we argued that such a setup corresponds to a quantum battery~\cite{2024arXiv240701832H}. The inequality~\eqref{eq:formula} is still valid for the quantity~\eqref{eq:delta}, where $\sigma$ and $\rho$ are replaced with $\rho_\text{QB}$ and $\rho_{\max}$, respectively. Therefore one cannot charge energy beyond the limit related to the fidelity between $\rho_\text{QB}$ and $\rho_{\max}$. One can also confirm  similar diagrams (Figs.~\ref{fig:teleportation},\ref{fig:fidelity}) using $\rho_\text{QB}$ instead of $\rho_\text{FC}$. 

This argument captures the core concepts of QET with one key difference. In QET, Bob can harness energy by simply allowing the system to evolve naturally over time. However, in the quantum battery context, he cannot gain additional energy since $\bra{E_{\max}}(U^\dagger H_{B}U)-H_B\ket{E_{\max}}$ is never greater than 0 for any unitary $U$, no matter how long he waits. So he must apply the feedback control operations to charge energy.

\section*{Acknowledgement}
I thank my friends at Stony Brook and BNL for many productive collaborations on quantum simulations over the years. In particular, I am grateful to Dmitri Kharzeev for his supervision during my postdoctoral period in his group, and to Fangcheng He, Hiroki Sukeno, and Tzu-Chieh Wei for useful discussions on the boundary conditions of the spin chains. I also appreciate Adam Lowe for productive collaborations on QET, Robert-Jan Slager and Rajeev Singh for helpful discussions on Rydberg atoms through our previous work, and Masahiro Hotta for a fruitful collaboration on the quantum battery inspired by his seminal work on QET. My work was supported by the U.S. Department of Energy, Office of Science, National Quantum Information Science Research Centers, Co-design Center for Quantum Advantage (C2QA) under Contract No.DESC0012704. 

\section*{References}
\bibliographystyle{utphys}
\bibliography{ref}

\clearpage
\begin{widetext}
\begin{center}
\textbf{\large Supplementary Material}
\end{center}
\section{\label{sec:QET}Algorithm of Quantum Feedback-control}
\subsection{Preliminaries}
Here we describe the quantum algorithm of the quantum feedback-control that we investigated in the main text. Our system consists of two parties: Alice and Bob. Alice plays the role of a Maxwell's Demon, performing measurements and sending feedback to Bob, who applies control operations based on Alice's feedback. The system is outlined in Fig.~\eqref{fig:system}, where their local sites are indicated by $n_A$ for Alice and $n_B$ for Bob. We consider lattices under open boundary conditions (OBCs) and periodic boundary conditions (PBCs). In the main text, we focus on chiral Dirac fermions in (1+1) dimensions under OBCs. In this Supplemental Material, we present results from both PBCs and OBCs.  

\begin{figure}[H]
    \centering
    \includegraphics[width=0.3\linewidth]{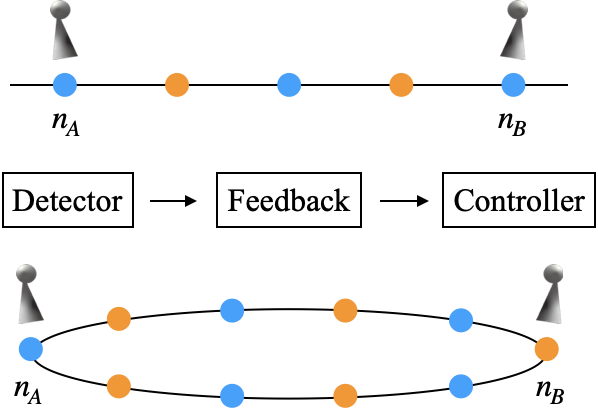}
    \caption{Outline of the setup, where Alice at $n_A$ is the detector and Bob at $n_B$ is the controller. They are spatially separated and communicate through classical channel.}
    \label{fig:system}
\end{figure}
\subsection{\label{sec:setp1}Step 1: Deposit Energy}
The following explanations in this section can be applied to a general spin chain. For simplicity, here we use the transverse Ising model of two qubits (minimal model of QET)~\cite{2010PhLA..374.3416H,2011arXiv1101.3954H}:  
\begin{equation}
    H=2k X_0X_1+h(Z_0+Z_1),~k,h>0. 
\end{equation}
We put Alice on $n_A=0$ and Bob on $n_B=1$, respectively. In the system, Alice and Bob interact directly through $X_0X_1$, so in order to ensure the non-triviality of the protocol, we use the following projective measurement operator for Alice: 
\begin{equation}
    P_0(b)=\frac{1}{2}(1+(-1)^b X_0).
\end{equation}
This operation does not affect Bob's energy since $[P_0,X_0X_1]=[P_0,Z_1]=0$, which allows us to define Bob's Hamiltonian as $H_B=2kX_0X_1+kZ_1$ and Alice's Hamiltonian as $H_A=hZ_0$. As we investigated in the main text, we do not restrict ourselves to energy and consider a general $2\times2$ Hermitian operator $O$ to explore \hyperref[prob:1]{Question \ref{prob:1}} from a differet perspective. We define $O_0=O\otimes I$ and $O_1=I\otimes O$, allowing us to ``teleport" a general observable from Alice to Bob. Such an $O$ can be written by a one-qubit unitary $U$ as $O=U+U^\dagger$. 

First, Alice measures her qubit in the ground state $\ket{gs}$ of the system. Then the net average amount she activates is evaluated by
\begin{equation}
    \langle \Delta O_{0}\rangle=\sum_{b\in\{0,1\}}\bra{gs}P_{0}(\mu)O_0P_{0}(b)\ket{gs}-\bra{gs}O_0\ket{gs}.
\end{equation}
Since her measurement device physically acts on the system, it also injects the energy $\langle\Delta E_0\rangle$ into the system. 

By extending the circuit provided in \cite{Ikeda:2023uni,Ikeda_Quantum_Energy_Teleportation_2023}, Alice's measurement can be implemented on a quantum circuit by finding a one-qubit unitary operator $U'$ such that $U'^\dagger ZU'=U$, meaning she measures her qubit in a general $U$-basis. In Fig~\ref{fig:random_Alice}, we show the net amount $\langle \Delta O_0\rangle$ and its absolute value $|\langle \Delta O_0\rangle|$ with respect to 300 random operators at each point $(k,h)$. Since $O$ is not necessarily her local Hamiltonian $H_A\propto Z$, the expectation value $\langle \Delta O_0\rangle$ could be positive or negative. While $\langle \Delta O_0\rangle$ seems to be completely random, it is interesting that the diagram of $|\langle \Delta O_0\rangle|$ is similar to the mean net energy $\langle \Delta E_A\rangle=\frac{h^2}{\sqrt{h^2+k^2}}$ she deposits (Fig.~\ref{fig:random_Alice}~(c)).

\begin{figure}[H]
    \subfigure[$\langle \Delta O_0\rangle$]{\includegraphics[width=0.32\textwidth]{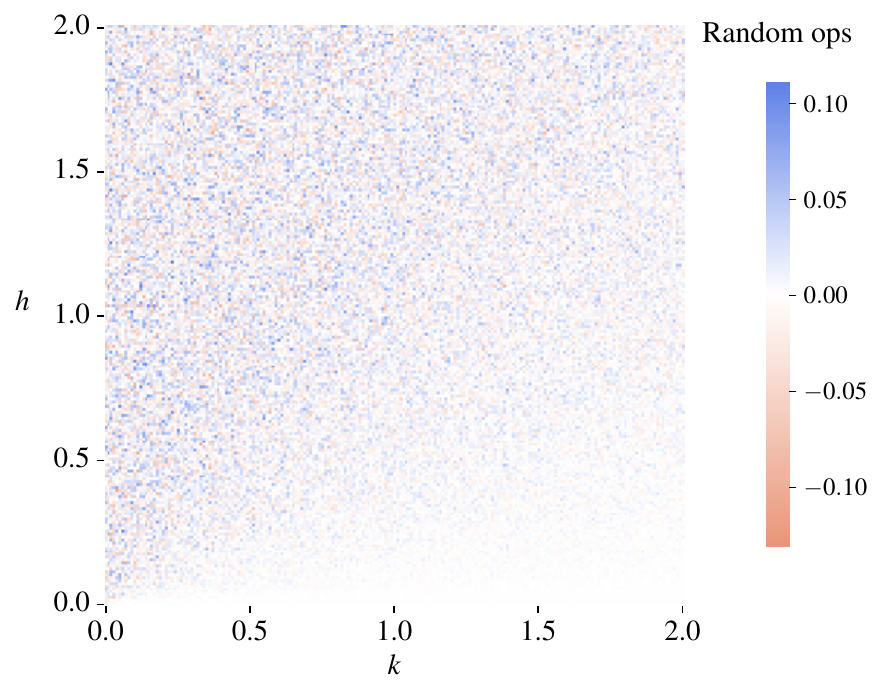}}
    \centering
    \subfigure[$|\langle \Delta O_0\rangle|$]
    {\includegraphics[width=0.32\textwidth]{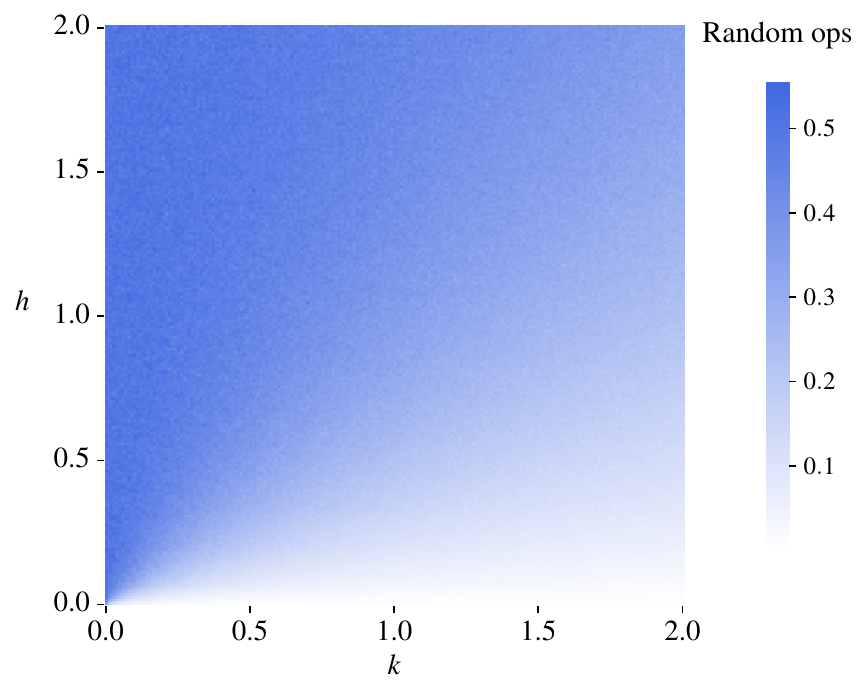}}
    \centering
    \subfigure[$\langle \Delta E_A\rangle$]
    {\includegraphics[width=0.3\textwidth]{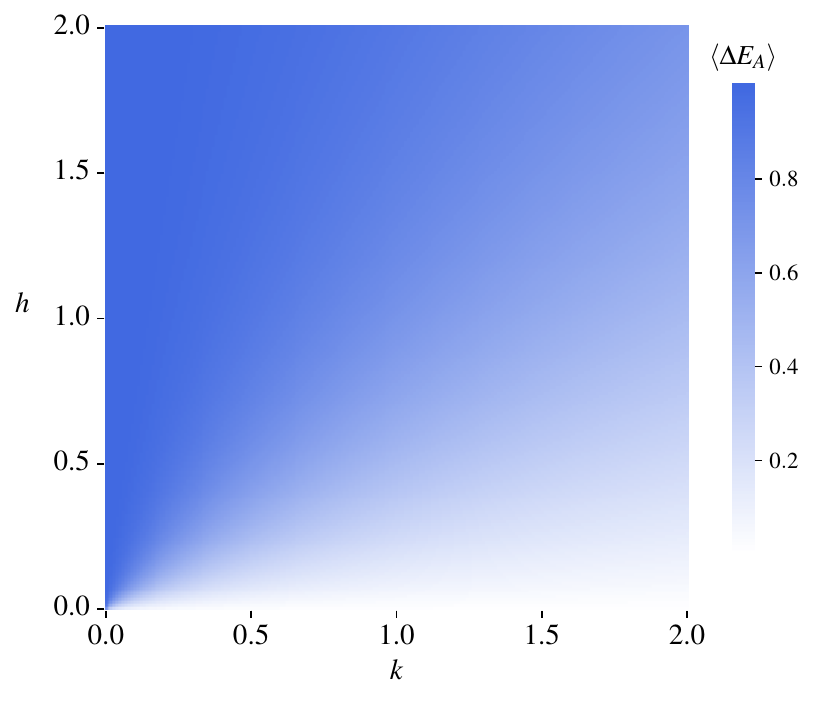}}
    \caption{Diagrams illustrating the activated amount by Alice's projective measurement. $O$ is a random 300 one-qubit Hermitian operator, sampled 300 times for each $(k,h)$. }
    \label{fig:random_Alice}
\end{figure}

\subsection{\label{sec:step2}Step 2: Teleport Energy and General Observables}
Bob performs control operations $U_1(b)$ based on Alice's feedback~$b\in\{0,1\}$:
\begin{equation}
    U_1(b)=\cos\phi I-i(-1)^b\sin\phi Y_1,
\end{equation}
where $\phi$ is defined by 
\begin{align}
    \cos(2\phi)=\frac{h^2+2k^2}{\sqrt{(h^2+2k^2)^2+h^2k^2}},~
    \sin(2\phi)=\frac{hk}{\sqrt{(h^2+2k^2)^2+h^2k^2}}.
\end{align}
In principle $\phi$ can be any value. Here $\phi$ is chosen so that the teleportation becomes optimal. More generally, for a Hamiltonian $H$ of a generic spin chain, such a $\phi$ is defined by 
\begin{align}
\label{eq:theta}
    \cos(2\phi)=\frac{\xi}{\sqrt{\xi^2+\eta^2}},~
    \sin(2\phi)=\frac{\eta}{\sqrt{\xi^2+\eta^2}},
\end{align}
where
\begin{align}
\label{eq:params}
\xi=\bra{gs}\sigma_{B}H\sigma_{B}\ket{gs},~\eta=\bra{gs}\sigma_{A}\dot{\sigma}_{B}\ket{gs},
\end{align}
with $\dot{\sigma}_{B}=i[H,\sigma_{B}]$. When $H$ is a locally-interacting Hamiltonian, it is important that the Hamiltonian satisfies $[H,\sigma_{B}]=[H_{B},\sigma_{B}]$. In the main text, we chose $\phi$ in this manner. To optimize the teleportation process, we also need to take $P_{n_A}$ and $U_{n_B}$ so that $\phi$ is not 0. 

The mixed quantum state $\rho_\text{FC}$ eq.\eqref{eq:rho_QET} is obtained statistically after the control operates $U_1(b)$ to $P_0(b)\ket{gs}$ are executed. The average net amount Bob can gain is 
\begin{equation}
\label{eq:QET}
    \langle \Delta O_1\rangle=\Tr[\rho_\text{FC}O_1]-\langle gs| O_1|gs\rangle. 
\end{equation}
Since Bob's measurement and control operations are done through his physical device, he can also gain the net energy $\langle \Delta E_B\rangle=\Tr[\rho_\text{FC}H_B]-\langle gs|H_B|gs\rangle$. 
      
We again repeated the protocol for 300 random one-qubit Hermitian operators $O$ in Fig.~\ref{fig:Bob_minimal}. It is intriguing that the phase diagram of the model is perfectly reconstructed by random operators. To illustrate this, compare Fig.~\ref{fig:Bob_minimal}(b) with Fig.~3 in \cite{Ikeda:2023uni}, which was evaluated based on the energy teleported to Bob's site. Fig.~\ref{fig:Bob_minimal}(c) shows the efficiency of the teleportation: the ratio between the output received by Bob and the input sent from Alice. 

\begin{figure}[H]
    \centering
    \subfigure[$\langle \Delta O_1\rangle$]{\includegraphics[width=0.3\textwidth]{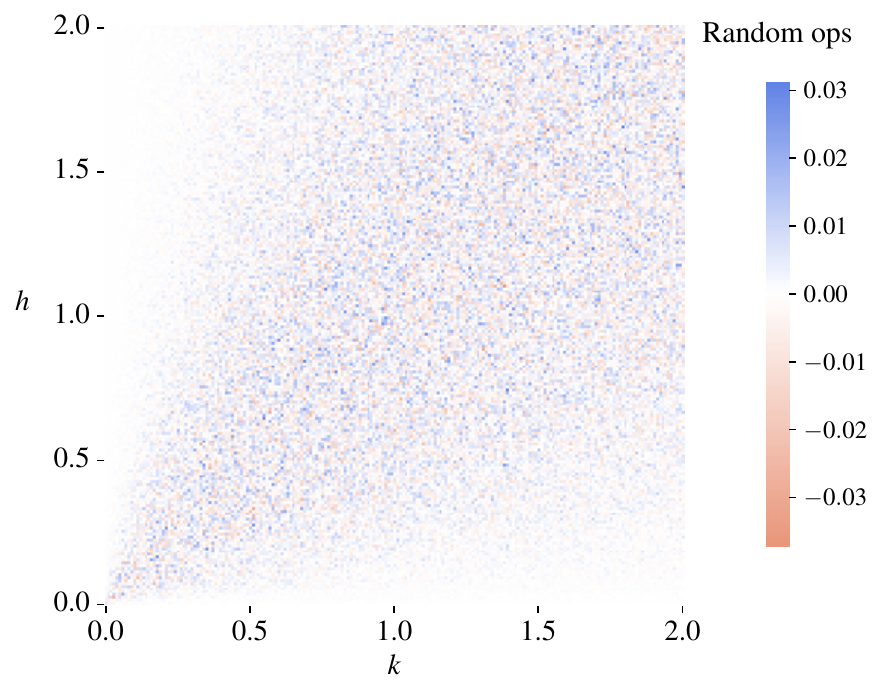}}
    \centering
    \subfigure[$|\langle \Delta O_1\rangle|$]
    {\includegraphics[width=0.3\textwidth]{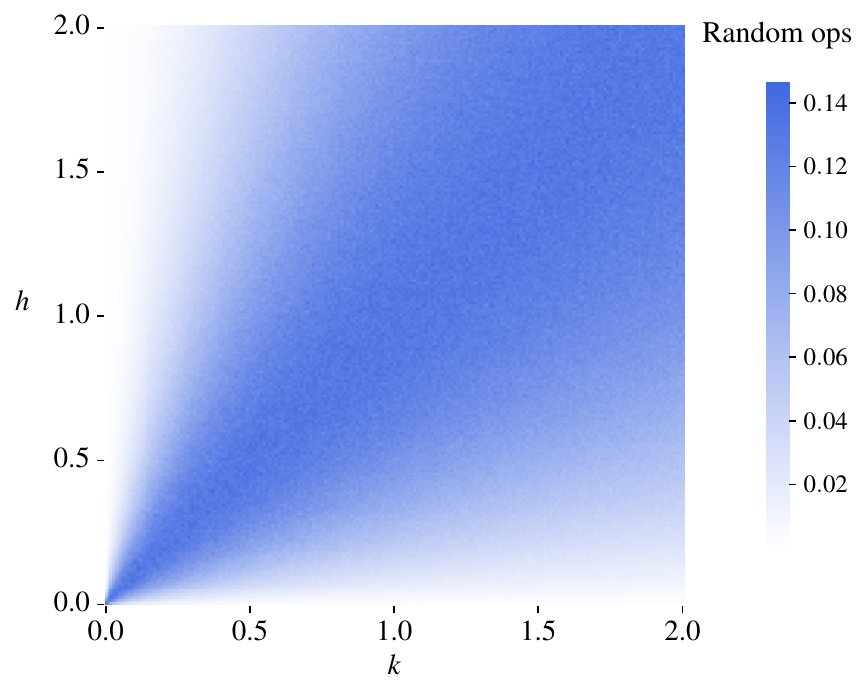}}
    \centering
    \subfigure[$|\langle \Delta O_1\rangle|/|\langle \Delta O_0\rangle|$]
    {\includegraphics[width=0.3\textwidth]{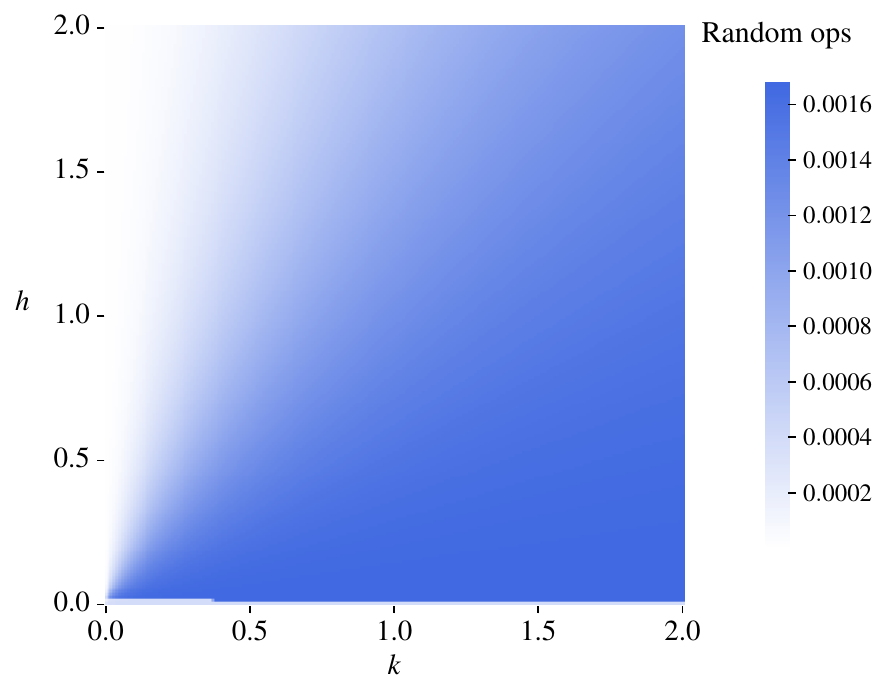}}
    \caption{Diagrams illustrating the teleported amount to Bob's site, evaluated by $\rho_\text{FC}$ and 300 random $O$s for each point.}
    \label{fig:Bob_minimal}
\end{figure}

\subsection{\label{sec:demon}Comment on the Maxwell Demon and QET}
The ground state of the minimal model is written as 
\begin{equation}
    \ket{gs}=\alpha\ket{00}+\beta\ket{11}.
\end{equation}
Before the measurement, her memory and the system is completely irrelevant so the mutual information is $I(X:A)=0$. 

Suppose Alice performs measurement using $P_0(b)=\frac{1}{2}(1+(-1)^b) X_0)$ to the first qubit. We write $\ket{gs}$ in $X$-basis as follows:
\begin{equation}
    \ket{gs}=\ket{+}\otimes\frac{\alpha\ket{0}+\beta\ket{1}}{\sqrt{2}}+\ket{-}\otimes\frac{\alpha\ket{0}-\beta\ket{1}}{\sqrt{2}}
\end{equation}
Therefore she can obtain $\ket{+}$ or $\ket{-}$ with probability $p_{\pm}=1/2$. The entropy is 
\begin{equation}
    S(X)=-p_{+}\log(p_{+})-p_{-}\log(p_{-})=\log2. 
\end{equation}
If there is no error in the measurement, the mutual information is $I(X:A)=S(X)=\log2$. 

Now let Alice make a feedback to the system, where Bob applies $U_1(b)=\cos\phi I+(-1)^{b}i\sin\phi Y$ to his qubit. After her measurement, the sate collapse either $\ket{+}$ or $\ket{-}$. Suppose she observes $b=0$. Then the state is projected to 
\begin{equation}
    \ket{\psi}=\ket{+}\otimes\frac{\alpha\ket{0}+\beta\ket{1}}{\sqrt{2}}
\end{equation}
with probability 1 and the feedback changes the state into 
\begin{equation}
    \ket{\psi'}=\ket{+}\otimes U_1(b=0)\left(\frac{\alpha\ket{0}+\beta\ket{1}}{\sqrt{2}}\right)
\end{equation}
with probability 1. Therefore the entropy of the system after the feedback is 0 and the mutual information is $I(X:A)=0$. 

As a result, the consumption of entropy $\Delta S(X)=-\log 2$ before and after the feedback is equal to the change in the mutual information $\Delta I(X:A)=-\log2$. In general, $\Delta S(X)\ge\Delta I(X:A)$ holds, which is called the second law of thermodynamics~\cite{PhysRevLett.100.080403,Parrondo2015}. In a thermal system with a heat bath at temperature $T$, work $W=-k_BT\Delta I(X:A)$ would be extracted.

One can also discuss this process with a quantum circuit where Alice's measurement is postponed to the end of the circuit~(see Fig.1 in \cite{Ikeda:2023uni}). In such a case, the control operation is performed first and the initial state is mapped into
\begin{equation}
    \ket{gs}\to\ket{\psi''}=\frac{1}{\sqrt{2}}\ket{+}\otimes U_1(b=0)\left(\frac{\alpha\ket{0}+\beta\ket{1}}{\sqrt{2}}\right)+\frac{1}{\sqrt{2}}\ket{-}\otimes U_1(b=1)\left(\frac{\alpha\ket{0}-\beta\ket{1}}{\sqrt{2}}\right).
\end{equation}
Again Alice observes $\ket{\pm}$ with equal probability $1/2$ and the system evolves in the same way.

\subsection{\label{sec:bound}Upper bound on teleportation: Entropy vs Fidelity}
Here we give a general remark on the upper bound of teleportation \eqref{eq:formula}. So far, the known best upper bound was given by the entanglement entropy between Alice's subsystem and its complement $S_{A\bar{A}}(\rho_0)$~\cite{PhysRevA.87.032313} in such a way that 
\begin{equation}
\label{eq:old_bound}
    \frac{E^2_{n_B}}{4\|H_{n_B}\|^2}\le S_{A\bar{A}}(\rho). 
\end{equation}
However, in general, the upper bound of $S_{A\bar{A}}(\rho_0)$ is determined by the area of $A$ (the number of qubits in $A$), the dimension of the system and the energy gap between the lowest energy and the 1st excited energy. Therefore $S_{A\bar{A}}$ can be larger than 1, depending on the systems and the parameters. In particular, when a system undergoes a quantum phase transition, it diverges at a critical point and becomes very large around it. Of course, the energy is finite around such a critical point, so the teleported energy is much lower than the entanglement entropy. In fact, the theoretically rigorous upper bound given by the fidelity $1-F(\rho_0,\rho_\text{FC})$ \eqref{eq:formula} guarantees that the ratio $E_{n_B}/4\|H_{n_B}\|$ does not exceed 1. A closer look at Figs.~\ref{fig:fidelity} and $\ref{fig:entropy_charge}$ clarifies that $1-F(\rho_0,\rho_\text{FC})$ can be greater or smaller than $S_{A\bar{A}}(\rho_0)$, depending on the points.   

Here we obtain the state-of-the-art upper bound of the teleported quantity. For a general observable $\mathcal{O}_{n_B}$ at Bob's site, we can still apply the discussion in~\cite{PhysRevA.87.032313}:
\begin{equation}
    \frac{|\langle\Delta\mathcal{O}_{n_B}\rangle-\langle\mathcal{O}_{n_B}\rangle|^2}{4\|\mathcal{O}_{n_B}\|^2}\le S_{A\bar{A}}(\rho_0),
\end{equation}
where $\langle\mathcal{O}_{n_B}\rangle=\sum_bp_A(b)\Tr[\rho_0U^\dagger_{n_B}(b)\mathcal{O}U_{n_B}(b)]$ with the probability $p_A(b)$ that $b$ is observed by Alice. When the sign of $\langle\Delta\mathcal{O}_{n_B}\rangle$ is opposite to the sign of $\langle\mathcal{O}_{n_B}\rangle$, we obtain a new upper bound: 
\begin{equation}
    \frac{|\langle\Delta\mathcal{O}_{n_B}\rangle|^2}{4\|\mathcal{O}_{n_B}\|^2}\le \min\{S_{A\bar{A}}(\rho_0),1-F(\rho_0,\rho_\text{FC})\}. 
\end{equation}
In particular, when $O_{n_B}$ is Bob's local Hamiltonian, the previous upper bound \eqref{eq:old_bound} is now updated as follows: 
\begin{equation}
\label{eq:new_bound}
    \frac{E^2_{n_B}}{4\|H_{n_B}\|^2}\le \min\{S_{A\bar{A}}(\rho_0),1-F(\rho_0,\rho_\text{FC})\}. 
\end{equation}

\section{\label{sec:Ham}More on teleportation in the (1+1)d chiral Dirac model}
\subsection{Hamiltonian of the model}
Here we give the qubit representation of the Hamiltonian that we address in the main text. This will be useful to perform experiments with a real quantum hardware. The action of the model in $(1+1)$-dimensional Minkowski space is
\begin{align}
 S = \int d^2x\mathcal{L},
\end{align}
where $\mathcal{L}$ is the Lagrangian~\eqref{eq:model} of the (1+1)-dimensional chiral model. The corresponding Hamiltonian is given by
\begin{align}
\label{eq:H_Schwinger}
 H &= \int dx 
 \bar{\psi}(-\im \gamma^1D_1 +m+\mu_5\gamma^1)\psi,
\end{align}
with commutation relation $\{\psi(x),\bar{\psi}(y)\}=\gamma_0\delta(x-y)$.

To facilitate quantum computation, we map the theory~\eqref{eq:H_Schwinger} onto a spatial lattice. We introduce staggered fermions, $\chi_n$ and $\chi^\dag_n$~\cite{Kogut:1974ag, Susskind:1976jm}, where $n$ denotes an integer labeling a lattice site and $a$ represents the lattice spacing. A two-component Dirac fermion $\psi = (\psi^1, \psi^2)^T$ is mapped into a staggered fermion such that $\psi^1 (\psi^2) \rightarrow \chi_n / \sqrt{a}$ for odd (even) $n$. The resultant lattice Hamiltonian is
\begin{align}
\label{eq:lattice_total_ham2}
 H =
 -\frac{\im}{2a}\sum_{n=1}^{N-1}
 \big[\chi^\dag_{n+1}\chi_{n}-\chi^\dag_{n}\chi_{n+1}\big]
 \nonumber+ m\sum_{n=1}^{N} (-1)^n \chi^\dag_n\chi_n
 \nonumber+ \im\frac{\mu_5}{2}\sum_{n=1}^{N-1} \big[\chi^\dag_{n+1}\chi_{n}+\chi^\dag_{n}\chi_{n+1}\big].
\end{align}
The first term represents the kinetic term, which describes the hopping between neighboring sites. The second term is the fermion mass. The third term denotes the axial charge, equivalent to the electric current in (1+1) dimensions.

We employ the Jordan-Wigner transformation~\cite{Jordan:1928wi}
\begin{equation}
\label{eq:Jordan-Wigner}
 \chi_n = \frac{X_n-\im Y_n}{2}\prod_{i=1}^{n-1}(-\im Z_i),~
 \chi^\dag_n = \frac{X_n+\im Y_n}{2}\prod_{i=1}^{n-1}(\im Z_i),
\end{equation}
to convert the Hamiltonian to the spin Hamiltonian. 

Those correspondences are summarized in Tab.~\ref{tab:dic}.  Our definition of the Dirac matrices are as follows: $\gamma^0=Z,\gamma^1=iY,\gamma^5=\gamma^0\gamma^1=X$, where $X,Y,Z$ are Pauli matrices. 
\begin{table}[H]
\begin{center}
\begin{tabular}{c|c|c}\toprule
Dirac & Staggerd  & Pauli \\\hline
     $\overline{\psi}\psi$ & $\frac{(-1)^n}{a}\chi^\dagger_n\chi_n$ &  $\frac{(-1)^n}{2a}(Z_n+1)$ \\
     $\overline{\psi}\gamma_0\psi$ & $\frac{1}{a}\chi^\dagger_n\chi_n$ &  $\frac{1}{2a}(Z_n+1)$ \\
     $\overline{\psi}\gamma_1\psi$ & $\frac{1}{2a}(\chi^\dagger_n\chi_{n+1}+\chi^\dagger_{n+1}\chi_{n})$ &  $\frac{1}{4a}(X_nY_{n+1}-Y_nX_{n+1})$ \\
    $\overline{\psi}\gamma_5\psi$ & $\frac{(-1)^n}{2a}(\chi^\dagger_n\chi_{n+1}-\chi^\dagger_{n+1}\chi_{n})$ &  $-\frac{i(-1)^n}{4a}(X_nX_{n+1}+Y_nY_{n+1})$ \\
    $\overline{\psi}\gamma_1\partial_1\psi$ & $-\frac{1}{2a^2}(\chi^\dagger_n\chi_{n+1}-\chi^\dagger_{n+1}\chi_{n})$ &  $-\frac{i}{4a^2}(X_nX_{n+1}+Y_nY_{n+1})$ \\
\end{tabular}
\end{center}
    \caption{Relationship between the three distinct representations of fermionic fields.}
    \label{tab:dic}
\end{table}

The corresponding qubit Hamiltonians $H_\text{obc},H_\text{pbc}$ under the OBC and the PBC are written as 
\begin{equation}
\label{eq:Ham_obc}
H_\text{obc}=\frac{1}{4a}\sum_{n=1}^{N-1}(X_{n}X_{n+1}+Y_{n}Y_{n+1})+\frac{m}{2}\sum_{n=1}^{N}(-1)^nZ_n+\frac{\mu_5}{4}\sum_{n=1}^{N-1}(X_{n}Y_{n+1}-Y_{n}X_{n+1})
\end{equation}
\begin{align}
\begin{aligned}
\label{eq:Ham_pbc}
H_\text{pbc}=&\frac{1}{4a}\sum_{n=1}^{N-1}(X_{n}X_{n+1}+Y_{n}Y_{n+1})+\frac{m}{2}\sum_{n=1}^{N}(-1)^nZ_n+\frac{\mu_5}{4}\sum_{n=1}^{N-1}(X_{n}Y_{n+1}-Y_{n}X_{n+1})\\
&+\frac{(-1)^{\frac{N}{2}}}{4a}\left[\left(X_{N}X_1+Y_{N}Y_1\right)+\mu_5\left(X_{N}Y_1-Y_{N}X_1\right)\right]\prod_{n=2}^{N-1}Z_n. 
\end{aligned}
\end{align}

\subsection{Teleportation under OBC and PBC}
Under the OBC, it is easy to see that the commutator $[H,P_n]$ vanishes at $a\mu_5=1,m=0$. Here $P_n=\frac{1}{2}(1-(-1)^bX_n)$. Up to a multiplier $ib$, the commutation relation between $H$ and $P_n$ is written as 
\begin{equation}
    [H,P_n]\propto\frac{(1-a\mu_5)}{4a}Z_n(Y_{n+1}+Y_{n-1})-\frac{m(-1)^{n}}{2}Y_n. 
\end{equation}
Therefore the post-measurement state is also an eigenstate of the Hamiltonian at $a\mu_5=1,m=0$. Since we use $X_{n_A}$ for Alice's measurement and $Y_{n_B}$ for Bob's controlled operator, it is also straightforward that $[H,Y_{n_B}]=0$ at $a\mu_5=-1,m=0$, where no teleportation occurs under the OBC. 

Note that the Hamiltonian $H_\text{pbc}$ under PBC has all-to-all interactions designated by the string $\prod_{n=2}^{N-1}Z_n$, which generates the $\mathbb{Z}_2$ symmetry associated with the parity. Due to such global interactions, the meaning of the energy teleportation to a local site is not clear. Nevertheless, we can still discuss the current (charge) teleportations since they are always defined by bilocal (local) operators, respectively. 

Suppose Alice is located at $n_A=1$ on the periodic chain. First, it should be noted that Alice's local measurement $P_1$ does not affect the energy level at other sites, since it commutes with the global interactions:
\begin{equation}
    \left[P_1,\prod_{n=2}^{N-2}Z_n\right]=0. 
\end{equation}
Therefore the non-triviality of the argument is still maintained.

In Fig.~\ref{fig:teleportatio_PBC}~(a,b) we present the results of the quantum charge and current teleportation under the PBC. Fig.~\ref{fig:teleportatio_PBC}~(c) shows the fidelity between the feedback-controlled state $\rho_\text{FC}$ and the ground state $\rho_0$ under the PBC. 
Despite the differences in boundary conditions, all figures exhibit qualitatively similar characteristics as delineated by the OPB, as shown in the main text (Figs.~\ref{fig:teleportation} and \ref{fig:fidelity}).

It is important to observe that within the PBC, the interpretation of the QET lacks tangible clarity. This arises due to the Hamiltonian $ H_\text{pbc}$, encompassing an all-to-all interaction, which inherently fails to commute with Bob's operation $U_{n_B}$, as defined by $Y$.

Bob can select $Z$ for his $U_{n_B}$ operation, thereby ensuring commutativity with the product operator $\prod_{n=2}^{N-2}Z_n$. Nonetheless, it is important to note that if both Alice and Bob choose the $Z$ operator in their respective operations, the teleportation becomes invalid.

\begin{figure}[H]
\begin{minipage}{0.32\linewidth}
    \centering
    \includegraphics[width=\linewidth]{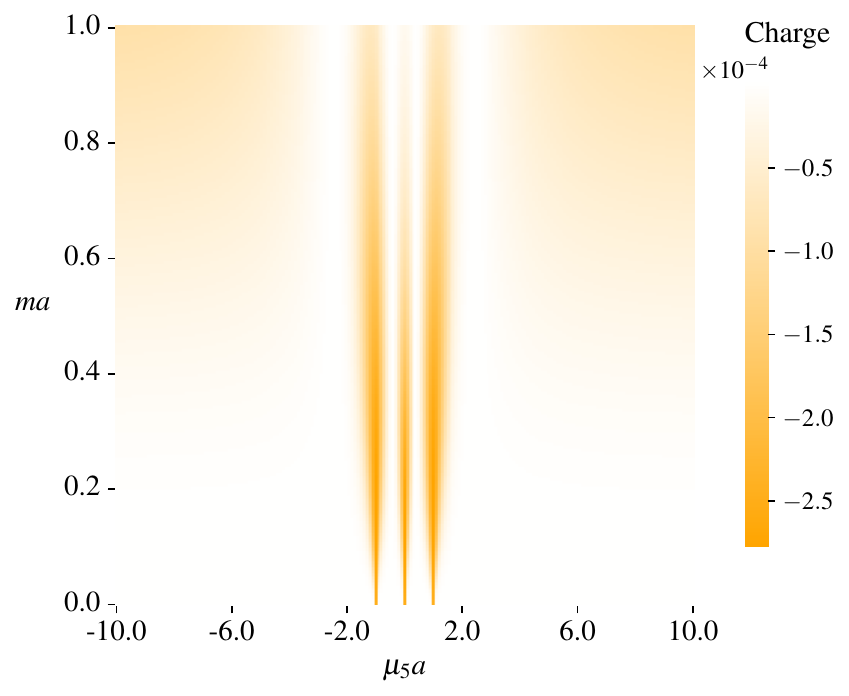}
\end{minipage}
\begin{minipage}{0.32\linewidth}
    \centering
    \includegraphics[width=\linewidth]{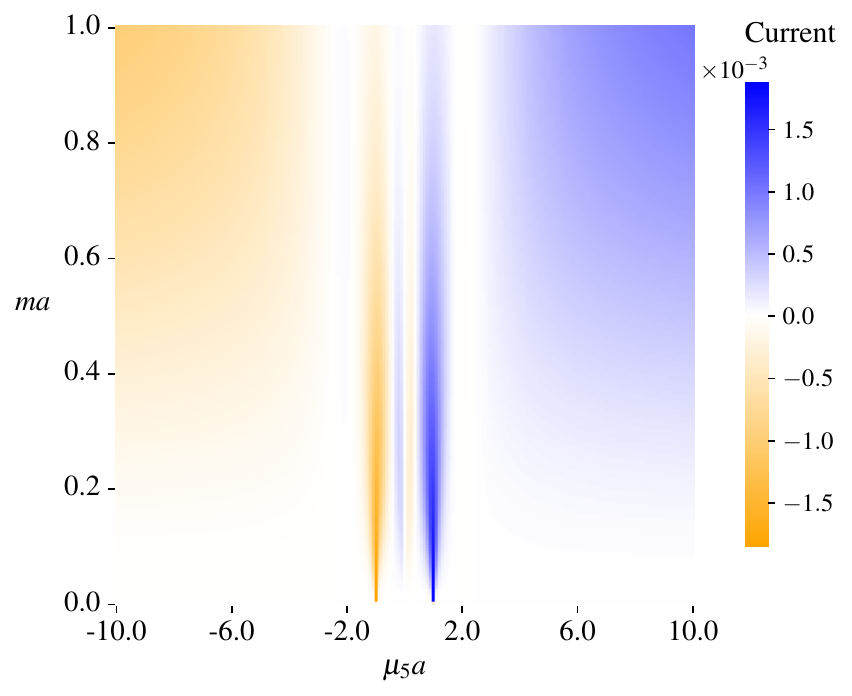}
\end{minipage}
\begin{minipage}{0.32\linewidth}
    \centering
    \includegraphics[width=\linewidth]{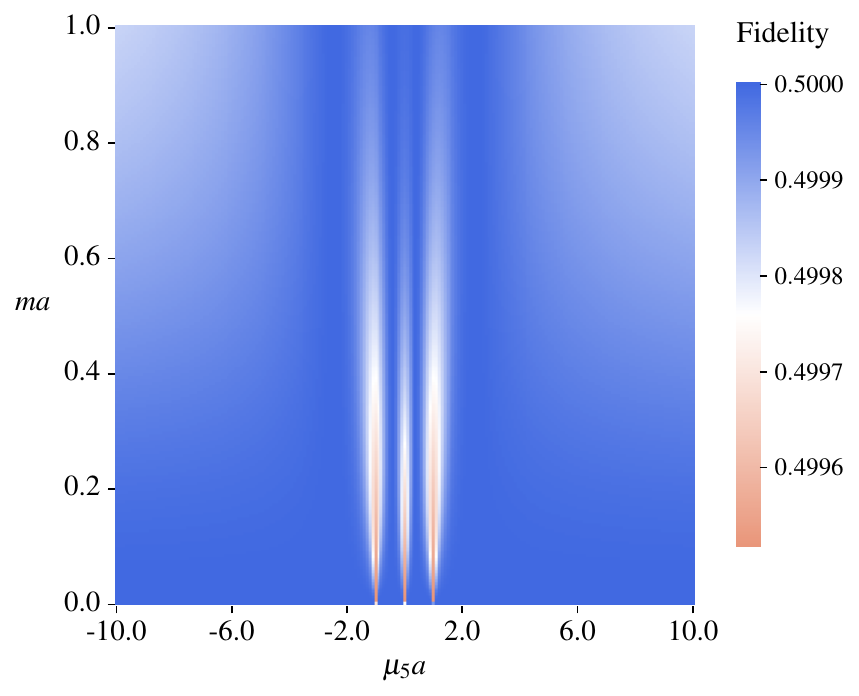}
\end{minipage}
    \caption{Charge and current under PBC ($N=8$).}
    \label{fig:teleportatio_PBC}
\end{figure}

\subsection{\label{sec:PBC_susceptibility}Charge susceptibility}
Here we consider the charge susceptibility for the PBC. The results within OBC are given in the main text (Fig.~\ref{fig:entropy_charge}). 

Fig.~\ref{fig:PBC_charge_susceptibility}~(a) presents a diagram derived from the ground state entanglement entropy between the half system and its complement, which can be interpreted as a phase diagram of the model. Meanwhile, Fig.~\ref{fig:PBC_charge_susceptibility}~(b) illustrates the diagram of the charge susceptibility as described by eq.~\eqref{eq:delta_susceptibility}. As highlighted in the main text, it is truly remarkable how the charge susceptibility precisely delineates the intricate structure of the phase diagram, including the peak and the local minimum of the diagram. As described in the main text, this is because the long-range correlations are created globally as exhibited in Fig.~\ref{fig:correlation_function}. This provide another evidence to our answer to \hyperref[prob:3]{Question \ref{prob:3}}. For the PBC, one can perform the same calculation and obtain a similar properties.   

In Fig.~\ref{fig:charge_susceptibility}, we plot the charge susceptibility as a function of chiral chemical potential under the OBC and the PBC. We also plot the differential of the susceptibility, $\partial\chi/\partial \mu_5$. Although the phase diagram and the teleportation are not directly related, it is noteworthy that $\partial\chi/\partial \mu_5$ qualitatively aligns with the local peak observed in the teleportation. This correspondence can be seen by comparing Figs.~\ref{fig:teleportation} and \ref{fig:charge_susceptibility}~(left), as well as Figs.~\ref{fig:PBC_charge_susceptibility} and \ref{fig:charge_susceptibility}~(middle).

\begin{figure}[H]
\begin{minipage}{0.49\linewidth}
    \centering
    \includegraphics[width=\linewidth]{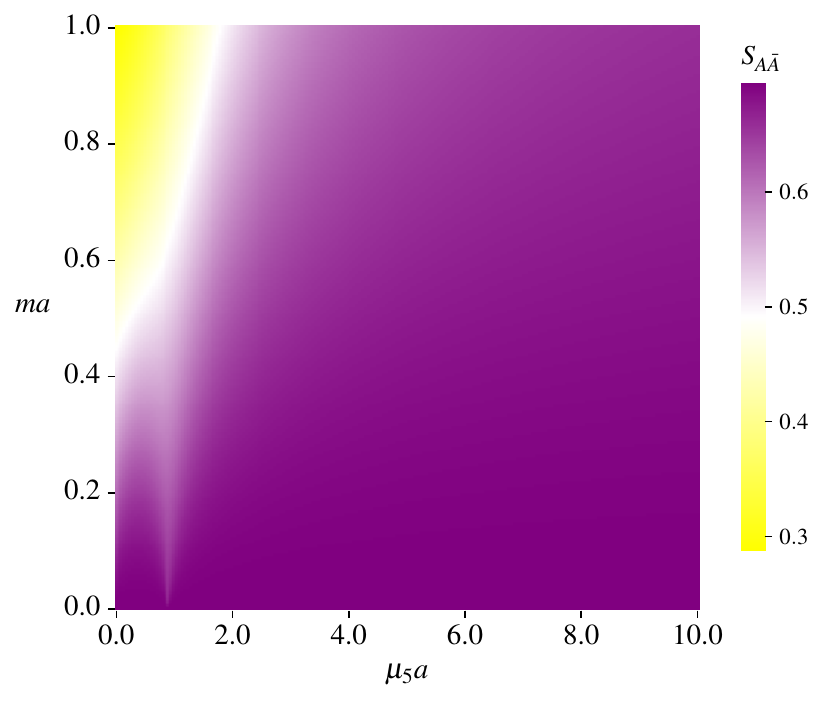}
\end{minipage}
\begin{minipage}{0.49\linewidth}
    \centering
    \includegraphics[width=\linewidth]{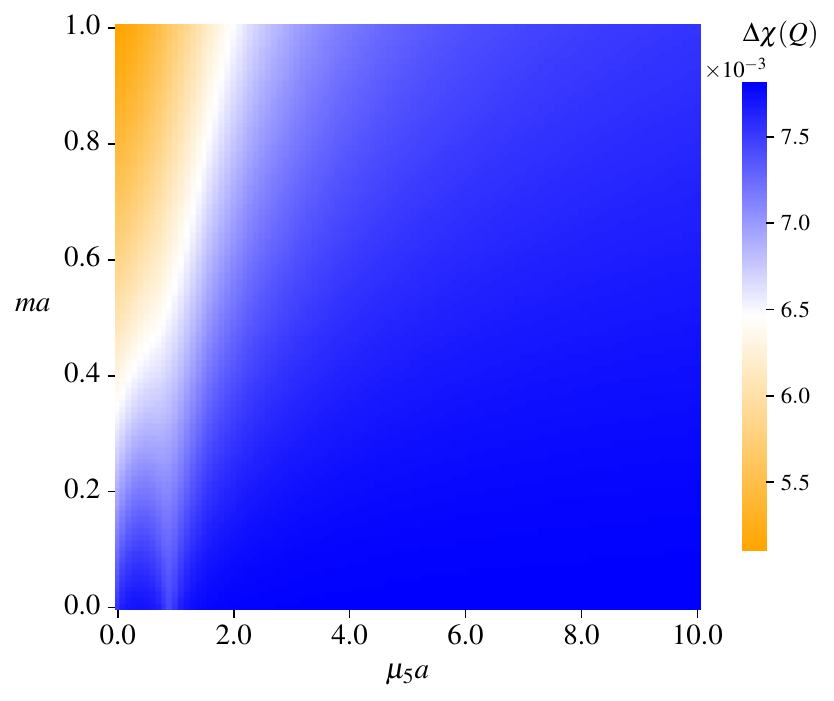}
\end{minipage}
    \caption{Left: the ground state entanglement entropy between Alice's system $A$ and its complement $\bar{A}$. Right: the charge susceptibility. $N=8$ under the PBC. }
    \label{fig:PBC_charge_susceptibility}
\end{figure}

\begin{figure}[H]
\begin{minipage}{0.32\linewidth}
    \centering
    \includegraphics[width=\linewidth]{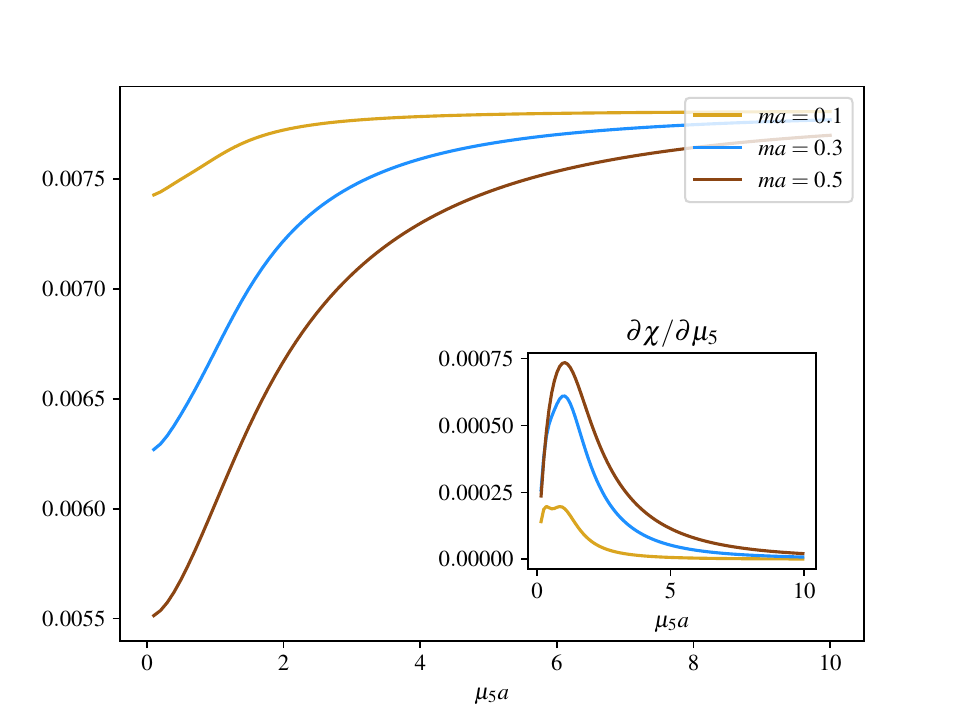}
\end{minipage}
\begin{minipage}{0.32\linewidth}
    \centering
    \includegraphics[width=\linewidth]{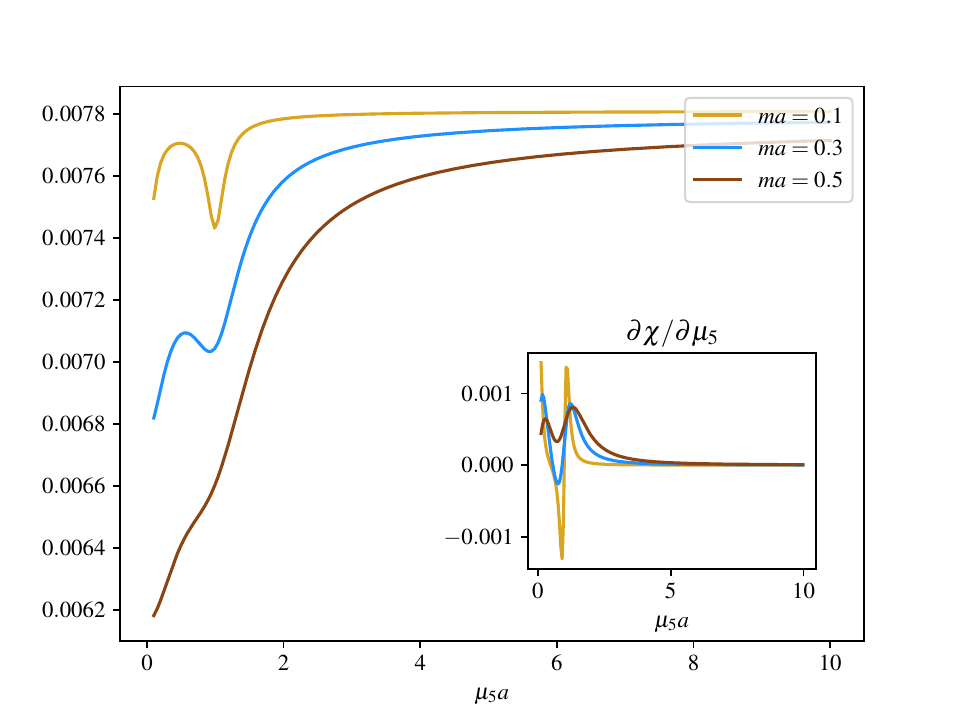}
\end{minipage}
\begin{minipage}{0.32\linewidth}
    \centering
    \includegraphics[width=\linewidth]{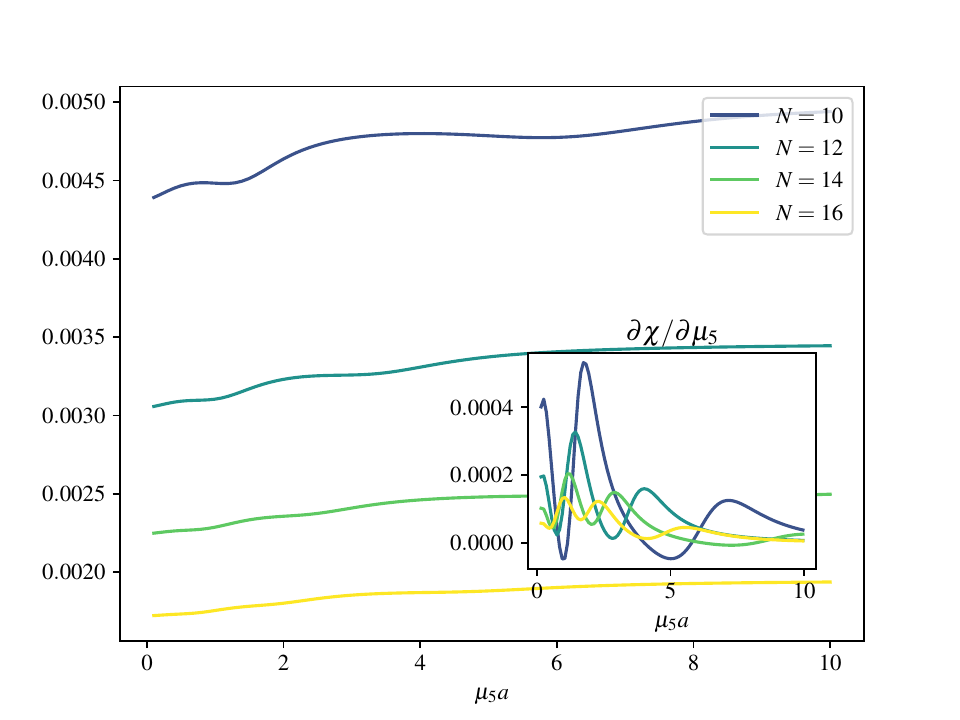}
\end{minipage}
\caption{Changes in the charge susceptibility induced by the feedback-control. From left to right: OBC $(N=8)$, PBC $(N=8)$, and $N$-dependence under PBC $(ma=0.3)$.}
    \label{fig:charge_susceptibility}
\end{figure}

\subsection{\label{sec:QQ}Real-time evolution of the charge-charge correlation function}
To investigate \hyperref[prob:3]{Question \ref{prob:3}} from yet another perspective, here we address the real-time evolution of the charge-charge correlation function~\eqref{eq:QQ-correlation} and the difference~\eqref{eq:diff_QQ-correlation}. A quantum circuit for implementing this simulation can be obtained by extending the circuit in \cite{Barata:2024bzk}, which is implemented in~\cite{code}. We combine the charges at even and odd sites $J^0_{2n}+J^0_{2n+1}$ in order to evaluate the physical charge. Bob is located at the center of the system, which is normalized to 0, and Alice is at $n_B-N/2$. 

Fig.~\ref{fig:time_evold_QQ}~(left) show the early time behavior of the charge-charge correlation function $\Delta\Pi^{00}_{0n_B}(t)$ between a generic site $n$ and Bob $n_B$. Fig.~\ref{fig:time_evold_QQ}~(right) is a 2d plot of the real-time evolution of $\Delta\Pi^{00}_{0n_B}(t)$, as a function of site $n$ and time $t$. We note that the light-conic propagation of the wave in Fig~\ref{fig:wave}~(b) is also confirmed here. The figure illustrate the fundamental property of the operator growth:
\begin{equation}
    J^0_{n_B}(t)=J^0_{n_B}+it[H,J^0_{n_B}]-\frac{t^2}{2}[[H,J^0_{n_B}],H]+\cdots.
\end{equation}

For a generic locally interacting Hamiltonian $H$ like ours, the $k$th-order nested commutator of $H$ with $J^0_n$ can result in a product of up to $k$ local operators, affecting a significant portion of the system. As a result, $[J_{n_B}(t), J_{n}]\neq 0$ and generally forms a large operator with substantial weight. The behavior of local operators has been widely explored in the contexts of quantum chaos, information scrambling, black holes, and many-body quantum teleportation~\cite{Roberts:2014isa,Hosur:2015ylk,PRXQuantum.5.010201}. 

Note that the ground state expectation value $\bra{gs}J^\mu_{n_B}(t)\ket{gs}$ is always constant and the total charge or current is conserved in the ground state. However, $J^\mu_{n_B}(t)$ oscillates when it follows quench dynamics. This can be also confirmed in QED in (1+1) dimensions~\cite{PhysRevD.108.074001,PhysRevResearch.2.023342}. As emphasized in the main text, Alice's measurement and Bob's control operation excite the energy level and induce a charge, therefore the system after those operations obeys the quench dynamics. For future work, it would be interesting to compute the out-of-time-ordered correlators (OTOCs) to further explore the effect of feedback on information scrambling.

\begin{figure}[H]
    \centering
    \includegraphics[width=0.45\linewidth]{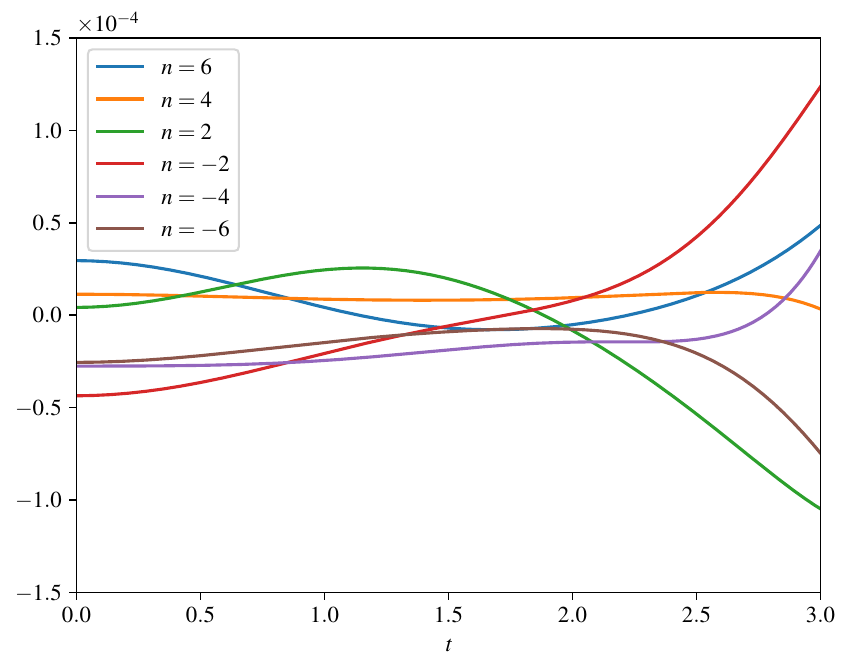}\centering
    \includegraphics[width=0.4\linewidth]{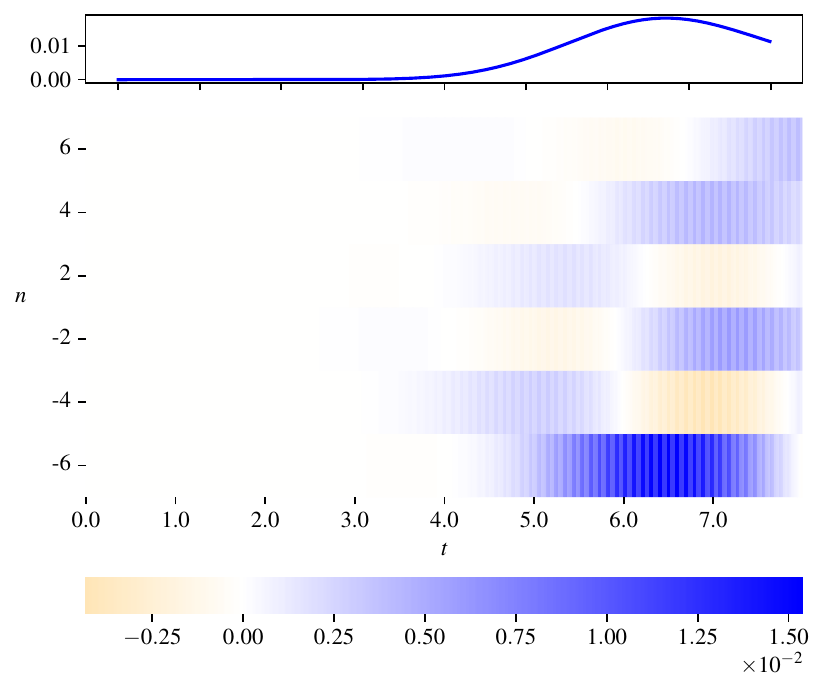}
    \caption{Real-time evolution of the charge-charge correlation function $\Delta\Pi^{00}_{0n_B}(t)$. The simulation was performed with parameters $ma=0.1,\mu=5a=1, N=16$ under the OBC. Similar plots can be obtained with the PBC.}
    \label{fig:time_evold_QQ}
\end{figure}

\section{(1+1)d Quantum Electrodynamics}
\subsection{Hamiltonian}
Here we consider the (1+1)d Quantum Electrodynamics (QED), called Schwinger model. The action of the massive Schwinger model with chiral chemical potential in $(1+1)$-dimensional Minkowski space is
\begin{align}
 S = \int\diff^2x\left[-\frac{1}{4} F^{\mu\nu} F_{\mu\nu} + \bar{\psi}(\im\slashed{D}-m-\mu_5\gamma^1)\psi\right],
\end{align}
with $\slashed{D}=\gamma^\mu(\p_\mu-\im gA_\mu)$.
Note that the gauge field $A_\mu$ and the coupling constant $g$ have mass dimensions 0 and 1, respectively. It will be convenient to use
the Hamiltonian formalism with the temporal gauge, $A_0=0$.

We map the model to a lattice using the staggered fermions and employ the Jordan-Wigner transformation~\eqref{eq:Jordan-Wigner}. For the details, see~\cite{PhysRevD.103.L071502}. The resulting Hamiltonian in the spin representation is expressed as follows:
\begin{equation}
\begin{split}
     H=&\frac{1}{4a}\sum_{n=1}^{N-1}(X_n X_{n+1}+Y_n Y_{n+1})+\frac{m}{2}\sum_{n=1}^N(-1)^n Z_n+\frac{\mu_5}{4}\sum_{n=1}^{N-1}(X_nY_{n+1}-Y_nX_{n+1})+\frac{a g^2}{2}\sum_{n=1}^NL_n^2.
 \end{split}
 \end{equation}
The first, second and third terms are the same as our previous model~\eqref{eq:Ham_obc}. The last term is the gauge kinetic term with the electric field given by
\begin{equation}
     L_n=\sum_{i=1}^{n}J^0_i=\sum_{i=1}^n\frac{Z_i+(-1)^i}{2}.
\end{equation}

\subsection{Quantum Current Teleportation}
Here we consider the current $J^1(x)$ of the Schwinger model. Although the model is not locally interacting, as it has the electric field $\sum_{n}L_n$, we can execute the protocol for teleporting current. It is crucial that the current operator acts only bilocally on the two sites. Thus, the concepts of current and charge teleportation remains valid.

The massive Schwinger model exhibits a critical point at $m/g \approx 0.33$, around which the teleported current displays an intriguing behavior, as illustrated in Fig.~\ref{fig:current_schwinger}. Our simulation was conducted with an $N=8$ lattice under the OBC. Here we used $\phi$ that is given by eq.~\eqref{eq:theta}, although the model is all-to-all interacting and thus $[H_B,\sigma_B]\neq[H,\sigma_B]$. 

\begin{figure}[H]
    \centering
    \includegraphics[width=0.45\linewidth]{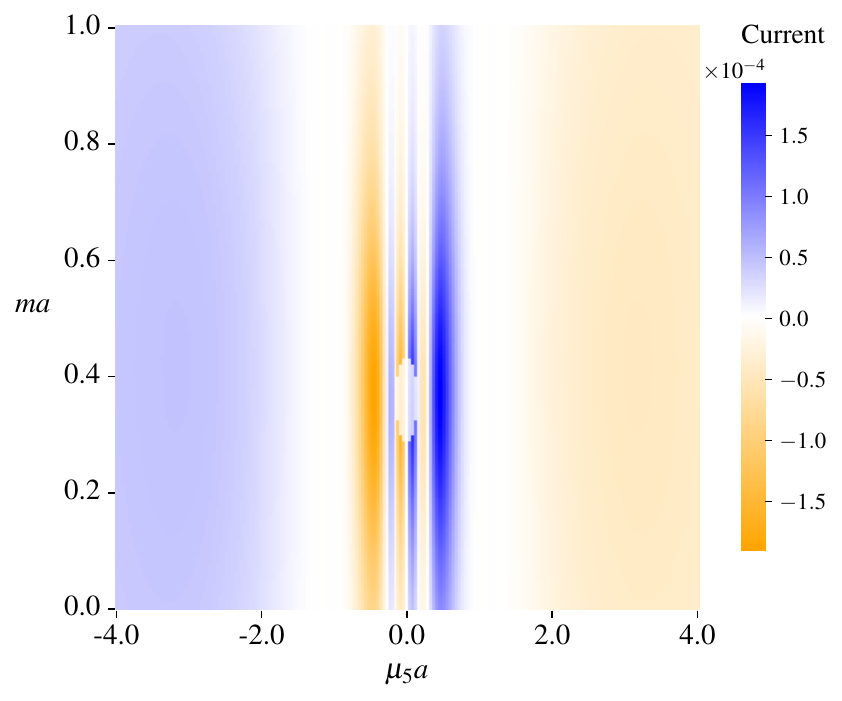}
    \caption{The net current teleported to Bob. $m/g\approx0.33$ corresponds to the critical point of the massive Schwinger model.}
    \label{fig:current_schwinger}
\end{figure}

\subsection{\label{sec:electric}Electric susceptibility}
In addition to the charge susceptibility, we explore electric susceptibility $\Delta \chi(L)$ using the electric field operator $L=\frac{1}{N-1}\sum_{n=1}^{N-1} L_n$, where $L_n=\sum_{j=1}^nJ^0_j$. The results are summarized in Fig.~\ref{fig:electric_susceptibility}, where the protocol was conducted with respect to the Schwinger model with a non-zero gauge field ($g\neq0$) and compared with the $g=0$ case, which is precisely the model we addressed in the main text~\eqref{eq:Ham_obc}. When $g\neq0$, the Hamiltonian is all-to-all interacting, therefore the meaning of energy teleportation is not clear. Nevertheless one can consider the feedback-control using the current or the charge operator, which is a bilocal or a local operator.  
\begin{figure}[H]
\begin{minipage}{0.40\linewidth}
    \centering
    \includegraphics[width=\linewidth]{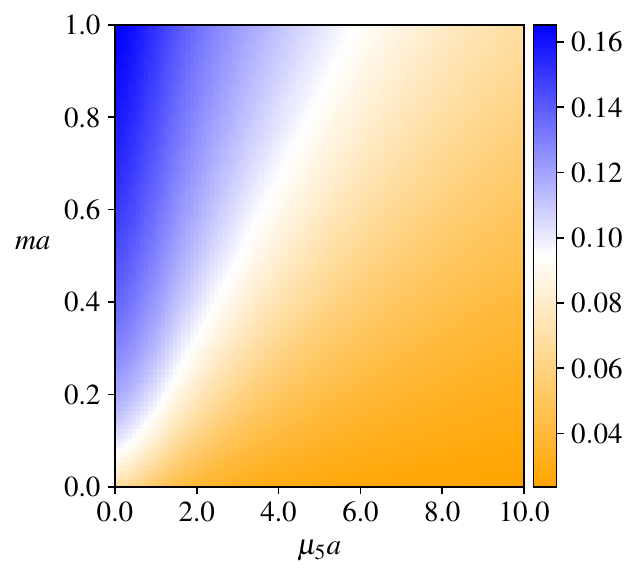}
\end{minipage}
\begin{minipage}{0.45\linewidth}
    \centering
    \includegraphics[width=\linewidth]{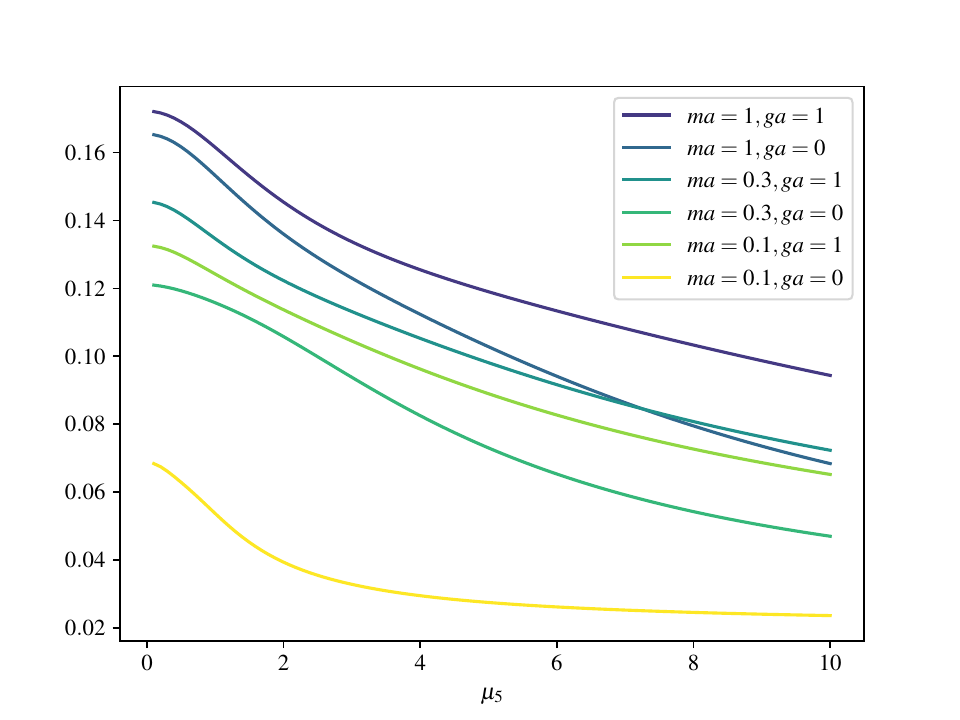}
\end{minipage}
    \caption{Difference in the electric susceptibility $\Delta\chi(L)$ of the Schwinger model. (OBC, $N=8$)}
    \label{fig:electric_susceptibility}
\end{figure}

\section{General extension of the teleportation to arbitrary coordinates}
As outlined in the main text, we provide an elaboration on the scenario where Alice and Bob execute the protocol on physical qubits or more general combinations of qubits. Specifically, they perform operations on neighboring qubits that consist of a Dirac fermion, which inherently possesses two components. For the sake of simplicity and manageable scale in the primary analysis, the operations were confined to only one component.

The choice of Alice and Bob's subsystems is entirely based on individual preferences, meaning that one has the freedom to choose where to place a detector and a controller in a given system. Let $S\subset\mathbb{Z}$ be the discrete set of all qubits in the system, $S_A\subset S$ be Alice's subsystem and $S_B\subset S$ be Bob's subsystem such that $S_A\cap S_B=\emptyset$. 

Then, the operators of Alice and Bob can be expressed in the most general form as
\begin{align}
    P_A(b)&=\frac{1}{2}\left(1-(-1)^b\prod_{n\in S_A}\sigma_n\right),\\
    U_B(b)&=\exp\left(-i2b\phi\prod_{n\in S_B}\sigma_n\right), 
\end{align}
where $\sigma_n$ is a one-qubit unitary and Hermitian operator at $n\in S$: $\sigma^\dagger_n=\sigma_n,\sigma_n^\dagger\sigma_n=I$. Using the operators, we can define the mixed state~\eqref{eq:rho_QET}. Note that $P_A(b)$ is a well-defined projector as it satisfies $P^\dagger_A(b)=P_A(b),P^2_A(b)=P_A(b)$ for all $b\in\{0,1\}$ and $P_A(0)+P_A(1)=I$.

A natural choice for labeling the subsystems will be $S=\{0,1,\cdots,N-1\},S_A=\{2i,2i+1\},S_B=\{2j,2j+1\}$, which allows one to apply operators to a Dirac fermion on a staggered lattice labelled by even and odd integers. 

We again use the chiral Hamiltonian~\eqref{eq:Ham_obc} and assume that Alice prefers to measure the chirality using $\gamma^5$ for her projector and that Bob uses $i\gamma^1$ for his control operation. In Fig.~\ref{fig:physical}, we present results of charge, current and energy teleported to Bob's coordinate $S_B=\{4,5\}$ from Alice's coordinate $S_A=\{0,1\}$ when they use $P_{A}(b)=\frac{1}{2}(1-(-1)^bX_{0} X_{1})$ and Bob's operation to $U_{B}(b)=\exp(-i2b\phi Y_{4}Y_{5})$, meaning that Alice measures the 1st physical qubit and Bob operates to the 3rd physical qubits in the lattice of 4 physical qubits.

\begin{figure}[H]
\begin{minipage}{0.32\linewidth}
    \centering
    \includegraphics[width=\linewidth]{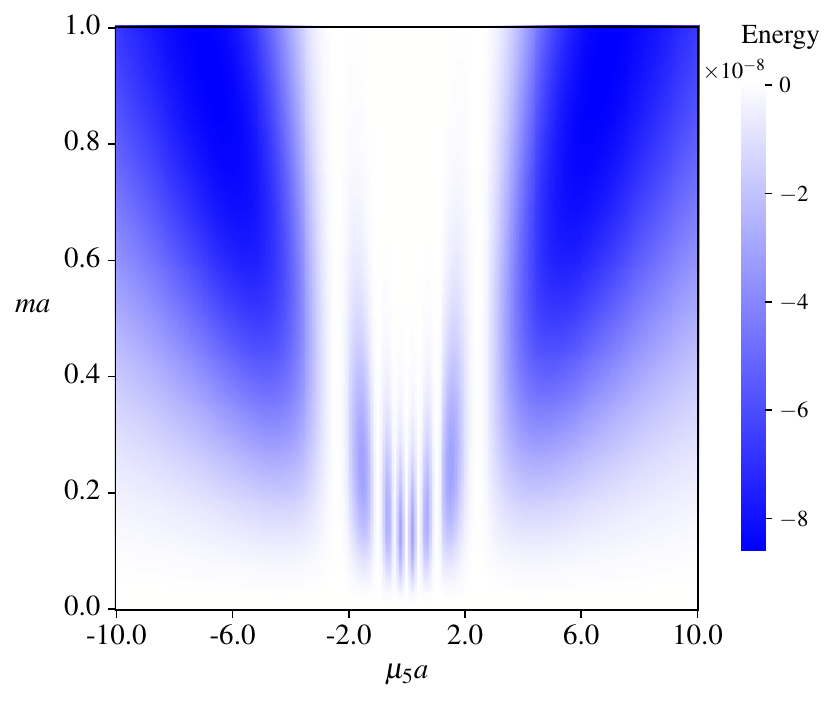}
\end{minipage}
\begin{minipage}{0.32\linewidth}
    \centering
    \includegraphics[width=\linewidth]{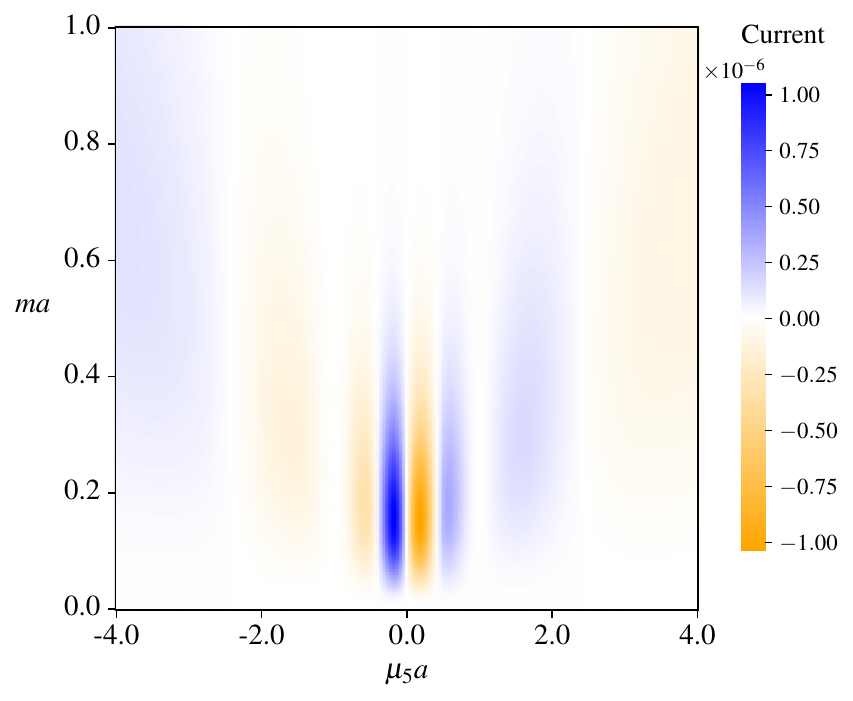}
\end{minipage}
\begin{minipage}{0.32\linewidth}
    \centering
    \includegraphics[width=\linewidth]{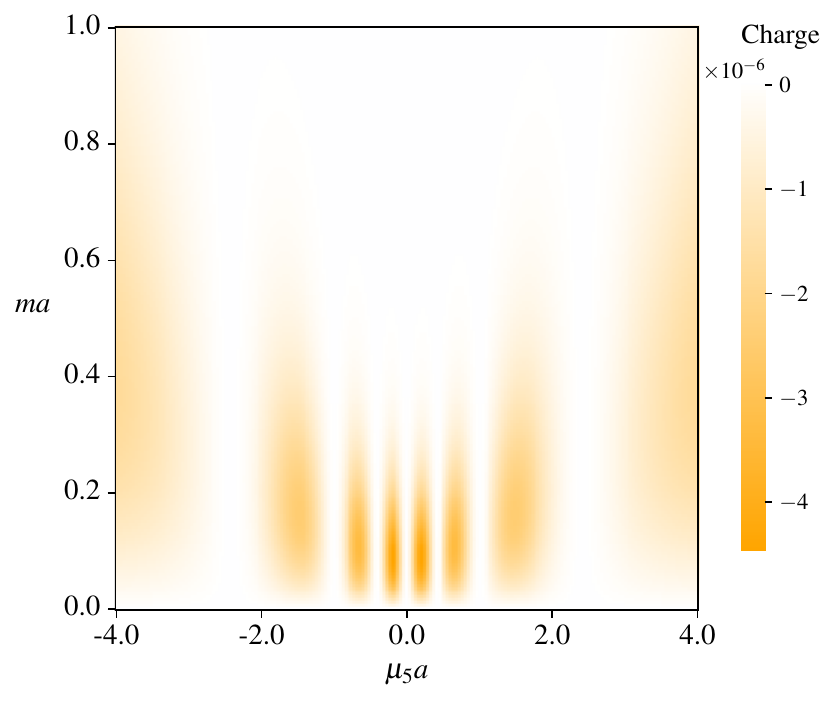}
\end{minipage}
    \caption{From left to right: Energy $\langle\Delta E_{n_B}\rangle$, electric current $\langle\Delta J^1_{n_B}\rangle$ and electric charge $\langle\Delta J^0_{n_B}\rangle$, computed with the lattice of $N=8$ qubits (4 physical qubits) under the OBC, using $P_{A}(b)=\frac{1}{2}(1-(-1)^bX_{0} X_{1})$ and $U_{B}(b)=\exp(-i2b\phi Y_{4}Y_{5})$ .}
    \label{fig:physical}
\end{figure}

\end{widetext}

\end{document}